\documentclass[a4paper,11pt]{article}
\pdfoutput=1

\usepackage{generic}

\let\address\affiliation

\usepackage[T1]{fontenc}
\usepackage[utf8]{inputenc}
\usepackage[english]{babel}
\usepackage[version=4]{mhchem}
\usepackage{subcaption}
\usepackage{multirow}
\usepackage{bm}
\usepackage{gensymb}
\usepackage{cite}
\usepackage{lineno}
\graphicspath{{combination_paper/}{./}}

\usepackage{accents}
\newcommand\brobor{\smash[b]{\raisebox{0.6\height}{\scalebox{0.5}{\tiny(}}{\mkern-1.5mu\scriptstyle-\mkern-1.5mu}\raisebox{0.6\height}{\scalebox{0.5}{\tiny)}}}}

\title{\boldmath Combined sensitivity of JUNO and KM3NeT/ORCA to the neutrino mass ordering}

\author[a]{S.~Aiello}
\author[ba,b]{A.~Albert}
\author[c]{M.~Alshamsi}
\author[d]{S. Alves Garre}
\author[e]{Z.~Aly}
\author[f,g]{A. Ambrosone}
\author[h]{F.~Ameli}
\author[i]{M.~Andre}
\author[j,\dag]{G.~Androulakis}
\author[k]{M.~Anghinolfi}
% ...
\address[a]{INFN, Sezione di Catania, Via Santa Sofia 64, Catania, 95123 Italy}
\address[b]{Universit{\'e}~de~Strasbourg,~CNRS,~IPHC~UMR~7178,~F-67000~Strasbourg,~France}
\address[c]{Universit{\'e} de Paris, CNRS, Astroparticule et Cosmologie, F-75013 Paris, France}
\address[d]{IFIC - Instituto de F{\'\i}sica Corpuscular (CSIC - Universitat de Val{\`e}ncia), c/Catedr{\'a}tico Jos{\'e} Beltr{\'a}n, 2, 46980 Paterna, Valencia, Spain}
\address[e]{Aix~Marseille~Univ,~CNRS/IN2P3,~CPPM,~Marseille,~France}
\address[f]{INFN, Sezione di Napoli, Complesso Universitario di Monte S. Angelo, Via Cintia ed. G, Napoli, 80126 Italy}
\address[g]{Universit{\`a} di Napoli ``Federico II'', Dip. Scienze Fisiche ``E. Pancini'', Complesso Universitario di Monte S. Angelo, Via Cintia ed. G, Napoli, 80126 Italy}
\address[h]{INFN, Sezione di Roma, Piazzale Aldo Moro 2, Roma, 00185 Italy}
\address[i]{Universitat Polit{\`e}cnica de Catalunya, Laboratori d'Aplicacions Bioac{\'u}stiques, Centre Tecnol{\`o}gic de Vilanova i la Geltr{\'u}, Avda. Rambla Exposici{\'o}, s/n, Vilanova i la Geltr{\'u}, 08800 Spain}
\address[j]{NCSR Demokritos, Institute of Nuclear and Particle Physics, Ag. Paraskevi Attikis, Athens, 15310 Greece}
\address[k]{INFN, Sezione di Genova, Via Dodecaneso 33, Genova, 16146 Italy}
% ...
\address[ba]{Universit{\'e} de Haute Alsace, rue des Fr{\`e}res Lumi{\`e}re, 68093 Mulhouse Cedex, France}
% ...

\author[]{\\(KM3NeT Collaboration)} 

\author[10]{\\S. Ahmad}
\author[4,*]{J. P. A. M. de André}
\author[4]{E. Baussan}
\author[13,9]{C. Bordereau}
\author[2]{A. Cabrera}
\author[13]{C. Cerna}
\author[8]{G. Donchenko}
\author[5]{E. A. Doroshkevich}
\author[4]{M. Dracos}
\author[13]{F. Druillole}
\author[13]{C. Jollet}
\author[4]{L. N. Kalousis}
\author[1]{P. Kampmann}
\author[8]{K. Kouzakov}
\author[5,8]{A. Lokhov}
\author[5]{B. K. Lubsandorzhiev}
\author[5]{S. B. Lubsandorzhiev}
\author[13]{A. Meregaglia}
\author[3]{L. Miramonti}
\author[13]{F. Perrot}
\author[4]{L. F. Piñeres Rico}
\author[8]{A. Popov}
\author[13]{R. Rasheed}
\author[11]{M. Settimo}
\author[8]{K. Stankevich}
\author[7,12]{H. Steiger}
\author[12]{M. R. Stock}
\author[8]{A. Studenikin}
\author[4]{A. Triossi}
\author[14]{W. Trzaska}
\author[8]{M. Vialkov}
\author[6]{B. Wonsak}
\author[4]{J. Wurtz}
\author[11]{F. Yermia}

\affiliation[1]{Forschungszentrum J\"{u}lich GmbH, Nuclear Physics Institute IKP-2, J\"{u}lich, Germany.}
\affiliation[2]{IJCLab, Universit\'{e} Paris-Saclay, CNRS/IN2P3, 91405 Orsay, France.}
\affiliation[3]{INFN Sezione di Milano and Dipartimento di Fisica dell Universit\`{a} di Milano, Milano, Italy.}
\affiliation[4]{IPHC, Universit\'{e} de Strasbourg, CNRS/IN2P3, F-67037 Strasbourg, France.}
\affiliation[5]{Institute for Nuclear Research of the Russian Academy of Sciences, Moscow, Russia.}
\affiliation[6]{Institute of Experimental Physics, University of Hamburg, Hamburg, Germany.}
\affiliation[7]{Institute of Physics, Johannes-Gutenberg Universit\"{a}t Mainz, Mainz, Germany.}
\affiliation[8]{Lomonosov Moscow State University, Moscow, Russia.}
\affiliation[9]{National United University, Miao-Li.}
\affiliation[10]{Pakistan Institute of Nuclear Science and Technology, Islamabad, Pakistan.}
\affiliation[11]{SUBATECH, Universit\'{e} de Nantes,  IMT Atlantique, CNRS-IN2P3, Nantes, France.}
\affiliation[12]{Technische Universit\"{a}t M\"{u}nchen, M\"{u}nchen, Germany.}
\affiliation[13]{Univ. Bordeaux, CNRS, CENBG, UMR 5797, F-33170 Gradignan, France.}
\affiliation[14]{University of Jyvaskyla, Department of Physics, Jyvaskyla, Finland.}

\author[]{\\(JUNO Collaboration members)}

\affiliation[\dag]{Deceased.}
\affiliation[\ddag]{also at Dipartimento di Fisica, INFN Sezione di Padova and Universit\`a di Padova, I-35131, Padova, Italy.}
\affiliation[*]{Corresponding authors.}

\emailAdd{jpandre@iphc.cnrs.fr}
\emailAdd{nchau@apc.in2p3.fr}
\emailAdd{elewyck@apc.in2p3.fr}

\abstract{
  This article presents the potential of a combined analysis of the JUNO and KM3NeT/ORCA
  experiments to determine the neutrino mass ordering.
  This combination is particularly interesting as it significantly boosts the potential
  of either detector, beyond simply adding their neutrino mass ordering sensitivities, by removing a degeneracy in the determination of $\Delta m_{31}^2$ between the two experiments
  when assuming the wrong ordering.
  The study is based on the latest projected performances for JUNO, and on  simulation tools using a full Monte Carlo approach to the KM3NeT/ORCA response with a careful assessment of
  its energy systematics.
  From this analysis, a $5\sigma$ determination of the neutrino mass ordering is expected
  after 6~years of joint data taking for any value of the oscillation parameters.
  This sensitivity would be
  achieved after only 2 years of joint data taking assuming the current global best-fit values for those
  parameters for normal ordering.
}

\begin{document}
%
%\linenumbers

\maketitle

\vspace*{1cm}
\section{Introduction}
\label{sec:intro}

The discovery of neutrino flavor  oscillations, implying that neutrinos are massive particles, is so far one of the few observational hints towards physics beyond the Standard Model. As such, it has potentially far-reaching implications in many aspects of fundamental physics and cosmology, from the matter-antimatter asymmetry in the Universe to the naturalness problem of elementary particles (see {\it e.g.} Refs.~\cite{Gonzalez-Garcia:2002bkq,Fukugita,Mohapatra}). Since the first conclusive observations of neutrino oscillations at the turn of the century~\cite{Fukuda:1998mi,Ahmad:2002jz,Eguchi:2002dm}, a variety of experiments targeting solar, reactor, atmospheric and accelerator neutrinos have achieved an increasingly precise determination of the parameters of the neutrino flavor mixing matrix~\cite{Zyla:2020zbs,Esteban:2020cvm, deSalas:2020pgw,Capozzi:2017ipn}. Despite this tremendous progress, some fundamental properties of neutrinos are yet to be determined, such as their absolute masses, whether they are Majorana particles and therefore are their own anti-particle, the existence and strength of CP-violation in the neutrino sector, and the ordering of the masses ($m_1$, $m_2$ and $m_3$) of the neutrino mass eigenstates (respectively, $\nu_1$, $\nu_2$ and $\nu_3$): either normal ordering (NO, $m_1 < m_2 < m_3$) or inverted ordering (IO, $m_3 < m_1 < m_2$). This last question is a prime experimental goal because its determination would have direct consequences on, {\it e.g.}, the measurement of leptonic CP violation in future long baseline experiments~\cite{Barger:2001yr} and the interpretation of results from planned experiments searching for neutrino-less double-beta decay to establish the Dirac vs.\@{} Majorana nature of neutrinos~\cite{Dolinski:2019nrj}.

The measurement of the neutrino mass ordering (NMO) is on the agenda of several ongoing neutrino experiments in the GeV energy domain that probe long-baseline \mbox{$\nu_\mu$ -- $\nu_e$} oscillations in Earth matter. Such experiments are  sensitive to the atmospheric mass splitting $\Delta m^2_{31}$~\cite{deSalas:2018bym}. However, none of these experiments alone, either  accelerator-based (such as T2K~\cite{Abe:2019ffx} or NO$\nu$A~\cite{Acero:2019ksn}) or using atmospheric neutrinos (such as Super-Kamiokande~\cite{Abe:2017aap} or  IceCube~\cite{Aartsen:2014yll}), has the capability to unambiguously resolve the NMO (in other words, the sign of  $\Delta m^2_{31}$) within the next few years. Even combining all available data, including those from reactor experiments, into global neutrino oscillation fits, has so far yielded only a mild preference for normal ordering, which has faded away again since the inclusion of the latest results of T2K~\cite{Abe:2021gky} and NO$\nu$A~\cite{Kolupaeva:2020pug}. Considering that a high-confidence ($>5\sigma$) determination of the NMO with the next-generation accelerator experiments DUNE~\cite{Acciarri:2015uup}, T2HK~\cite{Abe:2015zbg} and T2HKK~\cite{Abe:2016ero} is only envisaged for 2030 or beyond, alternative paths to the NMO measurement are being pursued on a shorter timescale.

JUNO~\cite{An:2015jdp,Djurcic:2015vqa,Abusleme:2021zrw} and KM3NeT/ORCA~\cite{Adrian-Martinez:2016fdl} are the two next-generation neutrino detectors aiming at addressing the NMO measurement within this decade. ORCA, the low-energy branch of the KM3NeT network of water Cherenkov neutrino telescopes, will determine the NMO by probing Earth matter effects on the atmospheric neutrino oscillations in the GeV energy range. JUNO is a medium-baseline ($\sim 53$~km) reactor neutrino experiment  that is sensitive to the NMO through the interplay between the fast oscillations driven by $\Delta m^2_{31}$ and $\Delta m^2_{32}$ in the $\bar{\nu}_e$ disappearance channel, where matter effects play only a small role~\cite{Li:2016txk}. As reactor $\bar{\nu}_e$ disappearance measurement is not affected by CP violation~\cite{An:2015jdp}, the JUNO measurement will be independent of the unknown CP violating phase $\delta_\mathrm{CP}$. Both detectors are currently under construction and target completion within the first half of this decade. The JUNO detector is planned to be completed in 2022, while ORCA is foreseen to be deployed incrementally until 2025, with 6 out of the total 115 detection lines already installed and taking data~\cite{ICRC_ORCA}.

Combining the data from the two experiments is not only motivated by their almost simultaneous timelines, but also by the gain in  sensitivity that arises from the complementarity of their approaches to the measurement of  the NMO. This boost essentially comes from the expected tension between ORCA and JUNO in the best fit of $\vert\Delta m^2_{31}\vert$ from the  $\bar\nu_e$ and $\nu_\mu$ disappearance channels when the assumed ordering is wrong. Provided that the measurement uncertainties in each experiment are small enough, the wrong mass ordering can be excluded with a high confidence level from the combination of the two datasets, even if the intrinsic NMO sensitivity of each experiment would not reach that level. This effect was first pointed out in relation with accelerator neutrino experiments ~\cite{ Nunokawa:2005nx, deGouvea:2005hk}, then reassessed in the context of the combination of a reactor experiment (Daya Bay II, now evolved into JUNO) and an atmospheric neutrino experiment (PINGU~\cite{TheIceCube-Gen2:2016cap}, a proposed low-energy extension of the IceCube neutrino telescope), showing that a strong boost in NMO sensitivity can indeed be reached with a combined fit~\cite{Blennow:2013vta}. The potential of this method was further explored in a combined study with JUNO and PINGU using detailed simulation tools for both experiments~\cite{Bezerra:2019dao}, leading to the same conclusion.

 In this paper, a complete study of the combined sensitivity of JUNO and ORCA to the NMO is presented, based on the same theoretical approach and using up-to-date detector configurations and expected performances. The main features and detection principles of the JUNO experiment are described in Sec.~\ref{sec:juno}, along with the standalone $\chi^2$ analysis used to determine the JUNO-only NMO sensitivity. The same is done for ORCA in Sec.~\ref{sec:orca},  based on the updated detector configuration, and  simulation and reconstruction tools used for the latest NMO sensitivity study following Ref.~\cite{Aiello:2021jfn}. In this case, special attention is paid to the treatment of  systematic uncertainties,  in particular with the introduction of a systematic error on the measured energy scale at the detector level (not considered in Ref.~\cite{Bezerra:2019dao} for PINGU). That systematic effect is shown to degrade the precision of the $\Delta m^2_{31}$ measurement of ORCA alone, thereby affecting the sensitivity of the combined study.

 The JUNO/ORCA combined $\chi^2$ analysis is presented in Sec.~\ref{sec:ana} for the baseline reactor configuration of JUNO, and the most realistic systematics treatment adopted for ORCA. The enhanced sensitivity achieved with the combined $\chi^2$ analysis over the simple sum of individual $\chi^2$ is also demonstrated.
 Sec.~\ref{sec:sens} presents further sensitivity studies exploring the impact of the energy resolution in JUNO and changing the number of reactor cores available towards a more optimistic scenario. The stability of the combined performance versus the true value of $\theta_{23}$ and $\Delta m^2_{31}$ is also addressed. The conclusions drawn from these results are presented in Sec.~\ref{sec:end}.

\section{Jiangmen Underground Neutrino Observatory (JUNO)}
\label{sec:juno}

The Jiangmen Underground Neutrino Observatory~\cite{An:2015jdp,Djurcic:2015vqa,Abusleme:2021zrw} (JUNO)
is a multipurpose experiment being built in the south of China.
Among its goals is the determination of the NMO
via the precise measurement of $\bar\nu_e$ from
the Yangjiang and the Taishan Nuclear Power Plants (NPP) located 53~km away from the detector.

The JUNO detector is divided into 3 parts: the Central Detector,
the Water Cherenkov Detector and the Top Tracker.
The Central Detector is composed of 20~kton of liquid scintillator placed in a
35.4~m diameter acrylic sphere.
Around this acrylic sphere, about 18k~20'' and 26k~3'' photomultiplier tubes (PMT)
monitor the liquid scintillator volume to detect neutrino interactions occurring inside,
in particular the Inverse Beta Decay (IBD) interactions produced by
$\bar\nu_e$ from the NPPs.
The IBD interactions are detected in the JUNO Central Detector via the prompt detection of the
scintillation light of the positron produced in the interaction and its subsequent annihilation,
along with
the delayed detection of a 2.2~MeV gamma-ray produced in the neutron capture on hydrogen and
subsequent de-excitation of the deuteron.
Due to the kinematics of the IBD, most of the available energy of the incident $\bar\nu_e$
is transferred to the positron.
Therefore, to do a precise measurement of neutrino oscillations,
a good energy resolution to measure the
visible energy of the prompt signal is critical, as will be discussed later.
The measured visible energy is smaller than the
incident $\bar\nu_e$ energy by about 0.8~MeV, due to the mass difference between the initial
and final particles ($-1.8$~MeV) and to the light emitted in the positron annihilation ($+1.0$~MeV).
The acrylic sphere is placed in the center of the Water Cherenkov Detector,
a cylindrical ultra-pure water pool
(44~m height, 43.5~m diameter), which serves to shield the Central Detector from
external radioactivity and to provide a veto for atmospheric muons and for muon-induced background
such as cosmogenic nuclei and fast neutrons.
The Water Cherenkov Detector and the Top Tracker, located on top of it to precisely track
atmospheric muons, compose the Veto System of JUNO.

In addition to the JUNO detector described above, the project also includes
the Taishan Antineutrino Observatory (JUNO-TAO) detector~\cite{Abusleme:2020bzt}.
This detector will be installed at a distance of 30~m from one of the Taishan's reactors to determine
the reactor $\bar\nu_e$ spectrum with a better energy resolution than JUNO,
effectively reducing the impact of possible unknown substructures in the
reactor neutrino spectra~\cite{Dwyer:2014eka} on the measurement of neutrino oscillations.

The precise distance of JUNO to each reactor core of the Yangjiang and Taishan NPPs, provided in
Ref.~\cite{An:2015jdp}, is used in this study rather than just the distance to the NPP complex.
In addition to them, the NPPs of Daya-Bay at 215~km and Huizhou at 265~km
will also contribute to the total number of detected reactor neutrinos.
However, given the much larger distance, the oscillation pattern will not be the same and these
neutrinos are part of JUNO's intrinsic background.
The Yangjiang NPP is already fully operational, with 6 reactor cores and a total
of 17.4~GW of thermal power, as is the Daya-Bay NPP, with a similar total thermal power.
The Taishan NPP has already 2 reactor cores operational out of the 4  initially foreseen,
totaling a thermal power of 9.2~GW. At present, it is unknown if the remaining 2 reactor cores,
which would bring another additional 9.2~GW of thermal power, will be built.
The Huizhou NPP is under construction and is expected to be ready by about
2025~\cite{Abusleme:2021zrw} with 17.4~GW thermal power.

\subsection{Modeling JUNO for this study}
\label{sec:juno:modeling}

For this study, the performance of JUNO closely follows that provided in Ref.~\cite{An:2015jdp}.
In particular, a 73\% IBD detection efficiency
and an energy resolution of $3\%/\sqrt{E/\text{MeV}}$
are assumed in the Central Detector which contains $1.5 \times 10^{33}$ target protons.
Given that the energy resolution is critical for the JUNO sensitivity, the impact of a
$\pm 0.5\%/\sqrt{E/\text{MeV}}$ change in energy resolution is discussed in
Sec.~\ref{sec:sens:juno}.
As in Ref.~\cite{An:2015jdp}, the nominal running time of the experiment is considered
to be 1000~effective days every 3~years.

The observed $\bar\nu_e$ spectrum in JUNO will be produced by the interplay between
the spectrum produced by the NPPs, the IBD cross-section~\cite{Vogel:1999zy},
and the neutrino oscillations which are to be measured.
To determine the spectrum produced by the NPPs, the
ILL\footnote{Institut Laue-Langevin}
$\bar\nu_e$ spectra~\cite{VonFeilitzsch:1982jw,Schreckenbach:1985ep,Hahn:1989zr}
are used given that the flux normalization is in better agreement with
previous data~\cite{Adey:2018qct}.
The fine structure of the spectrum will be precisely determined
independently using JUNO-TAO, as mentioned beforehand. Therefore, it is not
explicitly included in the spectrum shape.
The spectrum is calculated assuming a fission fragments content of
$\ce{^{235}U} : \ce{^{239}Pu} :\ce{^{238}U} : \ce{^{241}Pu} = 0.564 : 0.304 : 0.076 : 0.056$
which is similar to the one from Daya-Bay~\cite{An:2013zwz},
and using the fission energies for each of these isotopes from Ref.~\cite{Kopeikin:2004cn}.

In addition to reactor neutrinos, the IBD event selection contains some background
events.
The backgrounds considered in this analysis are taken from Ref.~\cite{An:2015jdp},
in terms of their rate, shape, and uncertainties.
The three dominant components of this background are cosmogenic events, geo-neutrinos, and
accidental coincidences mainly from radioactive background, with expected rates of about 1.6, 1.1, and 0.9~events per day, respectively.
As a reference, the Daya-Bay and Huizhou NPPs are expected to yield a total of 4.6~events per day
in JUNO
while the Taishan and Yangjiang NPPs are expected to produce a total of 54.3~events per day,
assuming the normal ordering world best-fit~\cite{Esteban:2018azc} oscillation parameters, considering all 4~Taishan NPP reactors operational.

A notable difference from Ref.~\cite{An:2015jdp} is our conservative choice to
consider, as baseline, only the 2 existing reactors in the Taishan NPP (referred to as ``JUNO 8~cores''
hereafter) rather than the
foreseen 4 reactors (referred to as ``JUNO 10~cores'' hereafter).
This reduces the total number of expected signal neutrinos by about 25\% in this study in comparison to Ref.~\cite{An:2015jdp}.
For completeness, the JUNO official baseline with a total of 4 Taishan reactors is also
considered and discussed in Sec.~\ref{sec:sens:juno}.
Although not yet completed, the Huizhou NPP  is considered to be active for the whole
duration of JUNO in both cases, even if it is possible that JUNO will start
before its completion.
This is again a conservative assumption given that the Huizhou NPP $\bar\nu_e$ are an
intrinsic background
to the neutrino oscillation measurements in JUNO.

\begin{figure}
  \begin{center}
    \includegraphics[width=.7\textwidth]{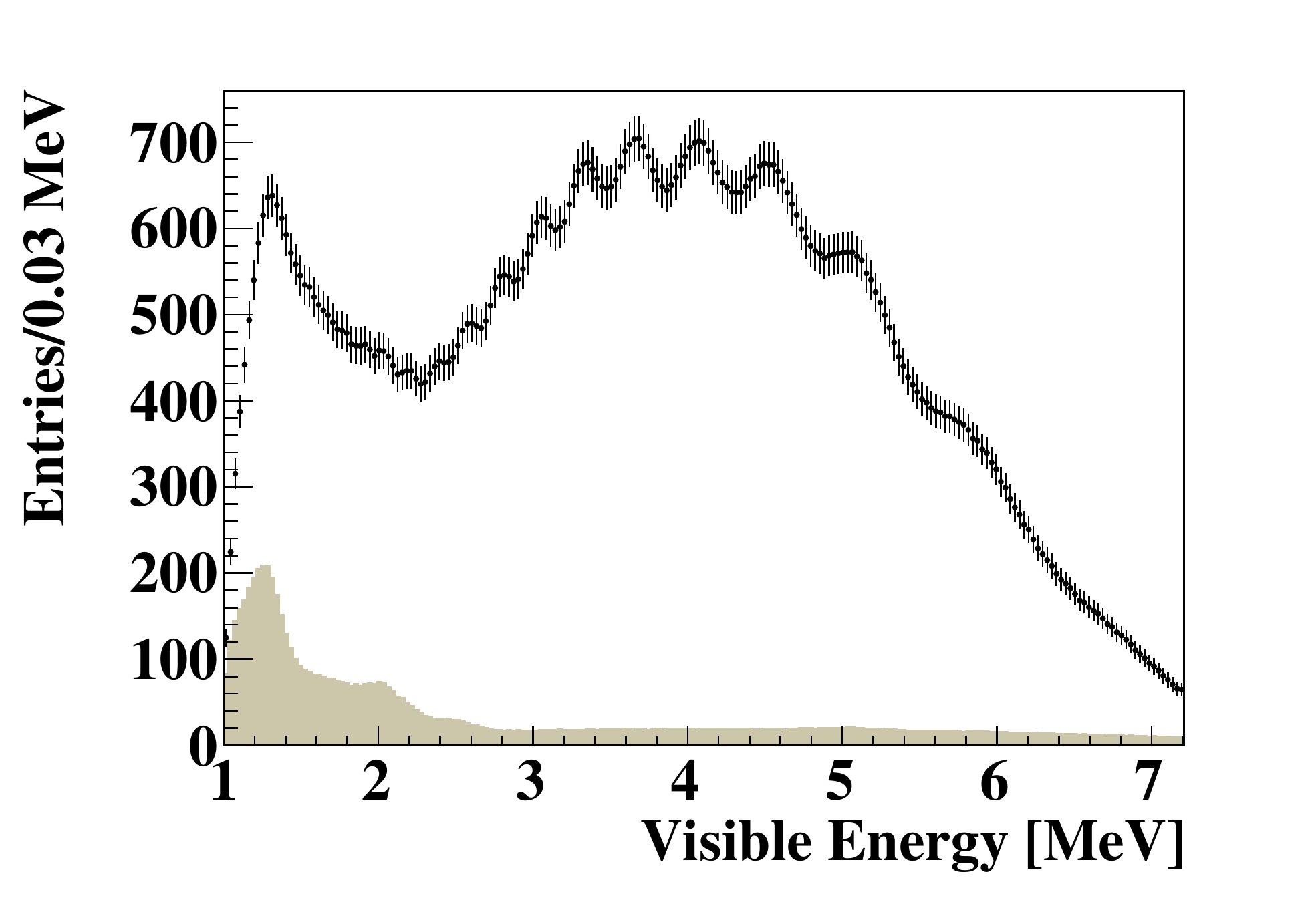}
    \caption{Expected event distribution for 6~years of data with JUNO 8~cores as a function
    of the visible energy of the prompt signal. The
    current world best-fit~\cite{Esteban:2018azc} oscillation parameters for normal ordering
    are assumed.
    The shaded region corresponds to the non-reactor neutrino background events.
    }
    \label{fig:juno:event_distribution}
  \end{center}
\end{figure}

The expected event distribution as a function of the visible energy is shown in
Fig.~\ref{fig:juno:event_distribution}, for 6~years of data with 8~cores and assuming the
best-fit oscillation parameters from Ref.~\cite{Esteban:2018azc} for normal ordering.
The events corresponding to the expected remaining non-reactor neutrino background
are also highlighted in the plot. They are concentrated in the lower energy part
of the measured spectrum where the energy resolution of JUNO is not sufficient to see
the rapid oscillation pattern.

\subsection{Sensitivity analysis}

In JUNO, the measurement of the neutrino oscillation parameters and, in particular,
of the neutrino mass ordering, is done by fitting the measured positron energy spectrum.
As shown in  Fig.~\ref{fig:juno:event_distribution},
this spectrum exhibits two notable features.
The first is
a slow oscillatory behavior due to $\theta_{12}$ and $\Delta m^2_{21}$,
which causes a large deficit in the number of events
in the whole energy range shown in Fig.~\ref{fig:juno:event_distribution} and that has a minimum at about 2.2~MeV.
The second is
a rapid oscillatory behavior due to $\theta_{13}$ and $\Delta m^2_{31}$ that starts
being visible in the figure at about 2~MeV, and for which several periods
are shown.
If inverted ordering is assumed rather than normal ordering in Fig.~\ref{fig:juno:event_distribution},
the position of the rapid oscillation maxima and minima would change, as the oscillation frequencies producing the pattern would be different~\cite{An:2015jdp}.
This is the signature
that JUNO will use to measure the NMO.
The sensitivity of JUNO to determine the NMO is calculated
using the $\chi^2$ difference between the data being fitted under the two ordering hypotheses. This strategy is also applied for the combined analysis.
While this procedure will be further detailed in Sec.~\ref{sec:ana}, it is useful to describe already at this point the $\chi^2$ function used
in JUNO.

It is also worth pointing out here that in this study no statistical fluctuations are
added to any of the simulated samples, and the expected statistical uncertainties at various
timelines are taken into account in the computed $\chi^2$ values.
This approach is commonly referred to as ``Asimov'' approach~\cite{Cowan:2010js}, and the
``Asimov'' term will be used in this paper to highlight that this approximation is being used.

For this analysis the measured JUNO visible energy spectrum is divided in $n=207$~bins of 0.03~MeV
between 1.00~MeV and 7.21~MeV.
The JUNO $\chi^2$ function used in this analysis has the following form:
\begin{equation}
\chi^2 = \bm{\Delta} \times \bm{\mathcal{M}}^{-1} \times \bm{\Delta}^T.
\label{eq:juno:chi2}
\end{equation}

In this expression, $\bm{\Delta}$ is a $1 \times n$ matrix whose content is the difference
between the observed and expected rates.
$\bm{\Delta}$ is defined as $\bm{\Delta} = \bm{D} - (\bm{S} + \bm{B})$,
where $\bm{D}$, $\bm{S}$, and $\bm{B}$ correspond, respectively, to the data, the signal prediction for a given
set of oscillation parameters, and the expected background.
$\bm{D}$ is given by the Asimov sample for the true value of the oscillation
parameters,
while $\bm{S}$ and $\bm{B}$ correspond to the expected signal given the test oscillation parameters
and the background.
$\bm{\Delta}^T$ is the transpose matrix of $\bm{\Delta}$.

The matrix $\bm{\mathcal{M}}$ whose inverse is present in Eq.~\eqref{eq:juno:chi2} is a
$n \times n$ covariance matrix.
This matrix is calculated as
$\bm{\mathcal{M}} = \bm{\mathcal{M}}_{stat} + \bm{\mathcal{M}}_{S} + \bm{\mathcal{M}}_{B}$.
$\bm{\mathcal{M}}_{stat}$ corresponds to
the statistical uncertainty in each bin from the total expected number of events ($\bm{S} + \bm{B}$).
$\bm{\mathcal{M}}_{S}$ and $\bm{\mathcal{M}}_{B}$ correspond, respectively, to the
covariance matrices of the signal and background as described in Ref.~\cite{An:2015jdp}.

This JUNO-only analysis was validated by comparing the NMO
sensitivity with previous results in Refs.~\cite{An:2015jdp,Bezerra:2019dao}.
Cross-checks have been performed using the same set of oscillation
parameters and reactor cores as in those references, showing agreement within 0.1--0.5 $\chi^2$ units.

% vim: spelllang=en

\section{Oscillation Research with Cosmics in the Abyss (ORCA)}
\label{sec:orca}

The KM3NeT Collaboration is currently building a set of  next-generation water Cherenkov telescopes in the depths of the Mediterranean Sea~\cite{Adrian-Martinez:2016fdl}. Two  tridimensional arrays of PMTs will be deployed: ARCA and ORCA (for \underline{A}stroparticle and \underline{O}scillation \underline{R}esearch with \underline{C}osmics in the \underline{A}byss, respectively). ARCA is a gigaton-scale detector which will mainly search for astrophysical neutrinos in the \mbox{TeV--PeV} energy range. ORCA, subject of this study, is a denser and smaller array (Mton-scale) optimized for oscillation physics with atmospheric neutrinos at energies above 1~GeV.

The ORCA detector will contain 115 detection units, each of them being a vertical line about 200~m long, anchored to the seabed and supporting 18 digital optical modules. These modules are glass spheres that contain 31~PMTs and related electronics. The array of detection units is arranged in a cylindrical shape with an average radius of 115~m and an average distance between the lines of 20~m. On each detection unit, the vertical spacing between the optical modules is around 9~m. The total instrumented volume of ORCA covers about $6.7 \times 10^6$ m$^3$, corresponding to 7.0 Mtons of seawater.

The ORCA PMTs detect the Cherenkov light induced by  charged particles originating in neutrino interactions in and around the detector. Such detectable interactions occur mainly through charged current (CC) and neutral current (NC) deep inelastic scattering processes off nucleons in water molecules~\cite{Formaggio:2013kya}. The  pattern and timing of the digitalized PMT output, or hits, recorded by the digital optical modules are used to identify neutrino events and reconstruct their energy and angular direction.
The topologies of neutrino-induced events in the  energy range of interest for ORCA can be separated into two broad classes. If the final state includes a sufficiently energetic muon, a track-like signature will be produced. This is the case for CC interactions of $\nu_\mu/\overline{\nu}_{\mu}$, and $\nu_\tau/\overline{\nu}_\tau$ with subsequent muonic decay of the tau lepton. Shower-like events correspond to all other interaction channels, where only hadronic and electromagnetic showers are produced. This includes CC interactions of $\nu_e/\overline{\nu}_e$ and $\nu_\tau/\overline{\nu}_\tau$ with non-muonic decays, as well as NC interactions of all flavors.

\subsection{Modeling the ORCA detector for this study}
 The analysis presented here is based on a detailed Monte Carlo (MC) simulation of the ORCA detector response to atmospheric neutrinos.
The generation of neutrino interactions in seawater in the energy range \mbox{1--100~GeV} is performed using gSeaGen~\cite{Aiello:2020kch}, a GENIE~\cite{Andreopoulos:2009rq}-based software developed within the KM3NeT Collaboration. All secondary particles  are tracked with the  package KM3Sim~\cite{Tsirigotis:2011zza} based on GEANT4~\cite{Agostinelli:2002hh} which simulates and propagates the photons induced by the Cherenkov effect, accounting also for light absorption and scattering,  and records the signals reaching the PMTs. The optical background due to radioactive decays of $^{40}$K naturally present in seawater is
simulated by adding uncorrelated (single-PMT) and correlated (inter-PMT) random noise hits, based on the rates measured with the first deployed detection units~\cite{Ageron:2019ksi}. The background of atmospheric muons is also simulated. The PMT response, readout, and triggering are simulated using KM3NeT custom software packages. The resulting trigger rate is about 54~kHz for noise events, 50~kHz for atmospheric muons and about 8~mHz for atmospheric neutrinos. The total simulated sample includes more than 15~years of atmospheric neutrinos,  1.4~days of noise events, and 14~days of atmospheric muons, which proves sufficient to probe the background contamination at a percent level~\cite{Aiello:2021jfn}.

The MC neutrino sample and event selection adopted in this study are identical to those  used in the latest ORCA NMO sensitivity study and are extensively described in  Ref.~\cite{Aiello:2021jfn}.
From the detected signals, the energy and direction of the events are reconstructed using dedicated algorithms developed for shower-like~\cite{Hofestaedt} and track-like~\cite{Quinn}  event topologies.

A set of preselection cuts are applied, requiring that events are well reconstructed, with an up-going direction (corresponding to a reconstructed zenith angle $\theta_\mathrm{reco} >90^\circ$), and obey certain criteria of containment in the detector instrumented volume. Events that pass the preselection cuts are then processed by a classification algorithm based on random decision forests for the determination of event topologies (or particle identification --- PID) and background rejection. Based on a single score scale $\eta$ from 0 to 1 provided by the classifier, the sample is then divided into 3 classes of events, also called PID classes: tracks ($0.7 < \eta \leq 1$), intermediate events ($0.3<\eta\leq 0.7$), and showers ($0\leq\eta\leq 0.3$). The two extreme PID classes provide a higher purity level for the genuine track-like and shower-like events.
As discussed in Ref.~\cite{Aiello:2021jfn}, the selection and background suppression cuts are sufficient to reduce pure noise events to a negligible rate and result in an atmospheric muon background contamination of only 3\%. The impact of such contamination is expected to be insignificant and atmospheric muons are not included in this study.

 The analysis relies on the computation of the expected energy and zenith angle $(E,\theta)$ distributions of atmospheric neutrino events for each PID class. Such distributions are obtained  with the SWIM package~\cite{Bourret:2018kug}, a KM3NeT analysis framework developed for calculating event distributions for a given hypothesis using a full MC approach to model the detector response.

 The incoming
atmospheric neutrino flux is taken from Ref.~\cite{Honda:2015fha}, for the Gran Sasso site without mountain over the detector, assuming minimum solar activity. The probabilities of neutrino flavor transitions along their path through the Earth are computed with the software OscProb~\cite{OscProb}, using a radial model of the Earth with 42 concentric shells of constant electron density, for which mass density values are fixed and follow the Preliminary Reference Earth  Model~\cite{Dziewonski:1981xy}. The rate of events interacting around the detector is computed for each interaction type $\nu_x \in \{\accentset{\brobor}\nu_e$~CC, $\accentset{\brobor}\nu_\mu$~CC,
$\accentset{\brobor}\nu_\tau$~CC, $\accentset{\brobor}\nu$~NC\} using neutrino-nucleon cross-sections weighted for water molecules as obtained with GENIE.

In order to obtain the expected event distribution as observed by the detector, the SWIM package uses a binned detector response matrix built from the MC sample that maps the events generated with interaction type $\nu_x$ and true variables $(E_\mathrm{true},\theta_\mathrm{true})$ into the corresponding reconstructed variables $(E_\mathrm{reco}, \theta_\mathrm{reco})$ and PID class (track, intermediate and shower). Given that there are 8 different interaction types and 3 different PID bins, the global response matrix is a collection of 24 4-dimensional matrices used for the transformation $(E_\mathrm{true},\theta_\mathrm{true}) \longrightarrow (E_\mathrm{reco}, \theta_\mathrm{reco})$.
These matrices are built using MC-generated events and the outcome of their processing through the reconstruction and classification algorithms, so that the ensemble of matrices account for detection and reconstruction efficiencies, misidentification probabilities and errors on reconstructed
variables (including all correlations). This approach is different from the one in Ref.~\cite{Aiello:2021jfn} which uses parametrized response functions obtained from the MC distributions.

While this full MC method ensures that all the information on the detector response, including potential correlations between parameters, is taken into account,
its accuracy depends on the size of the MC sample. To account for statistical fluctuations in the MC production that could affect the response functions used to build the matrix, the Beeston-Barlow light method~\cite{Barlow:1993dm} has been adopted, as described in the next section.

\begin{figure}[th]
    \centering
    \includegraphics[width=0.45\linewidth]{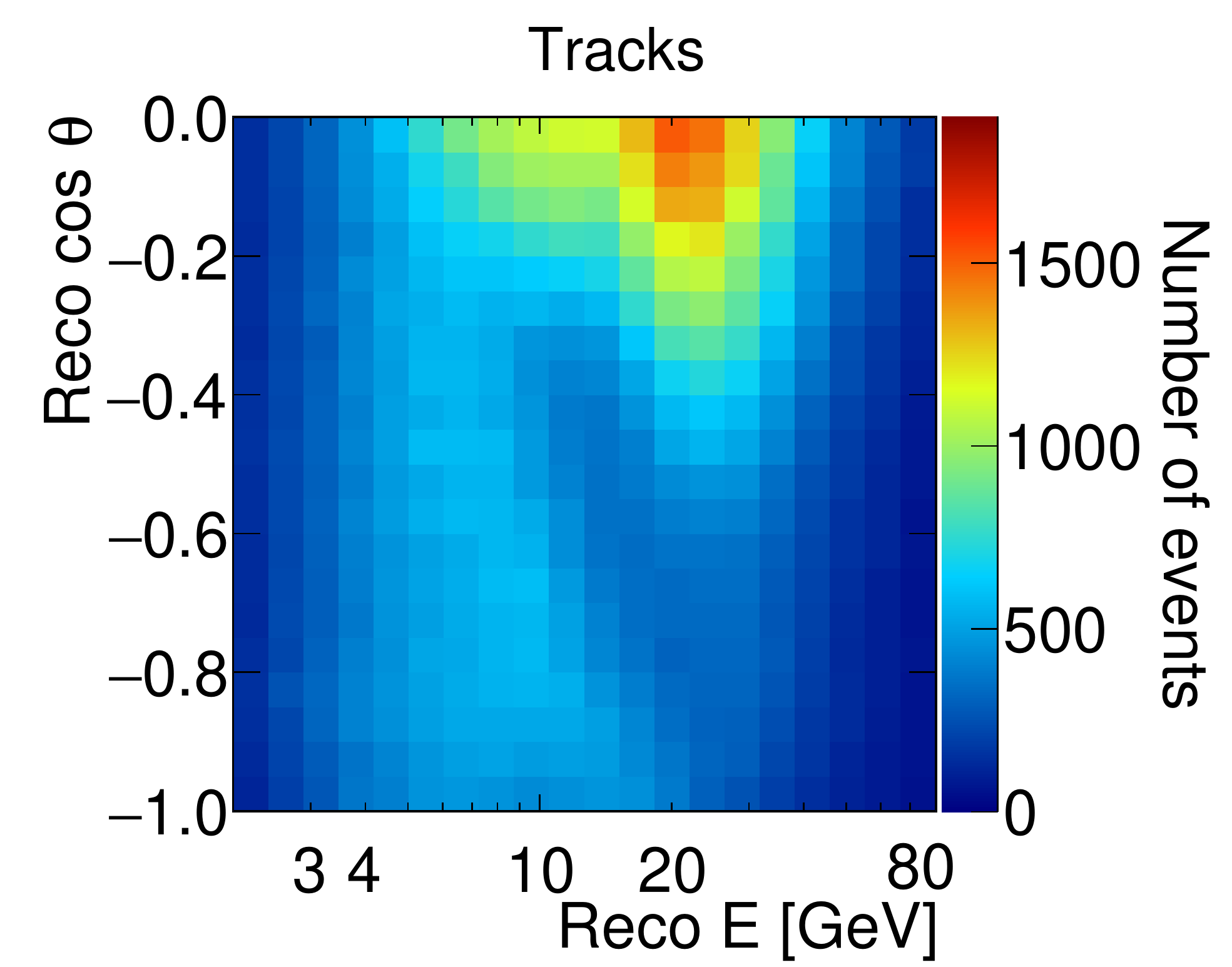}
    \includegraphics[width=0.45\linewidth]{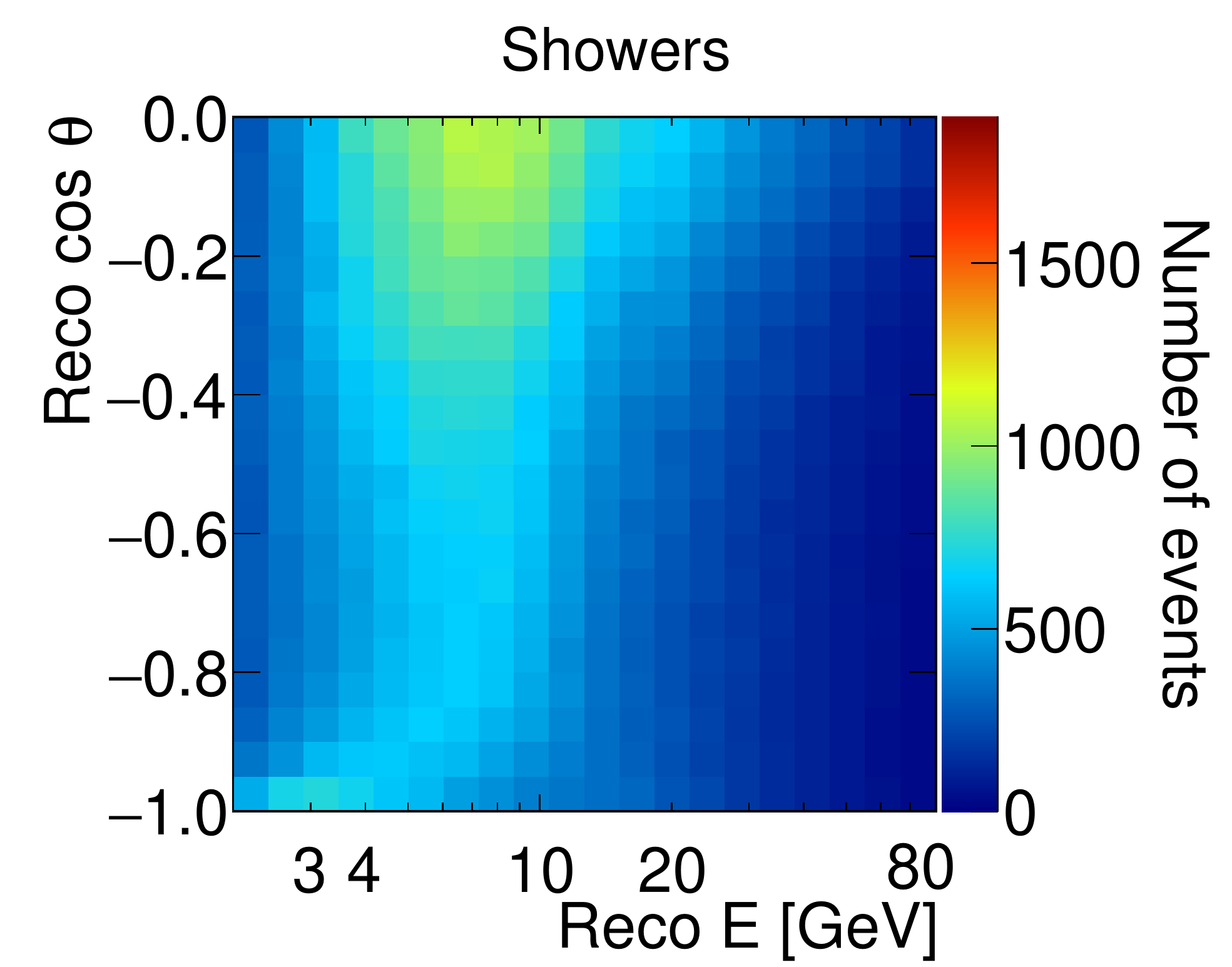}
    \includegraphics[width=0.45\linewidth]{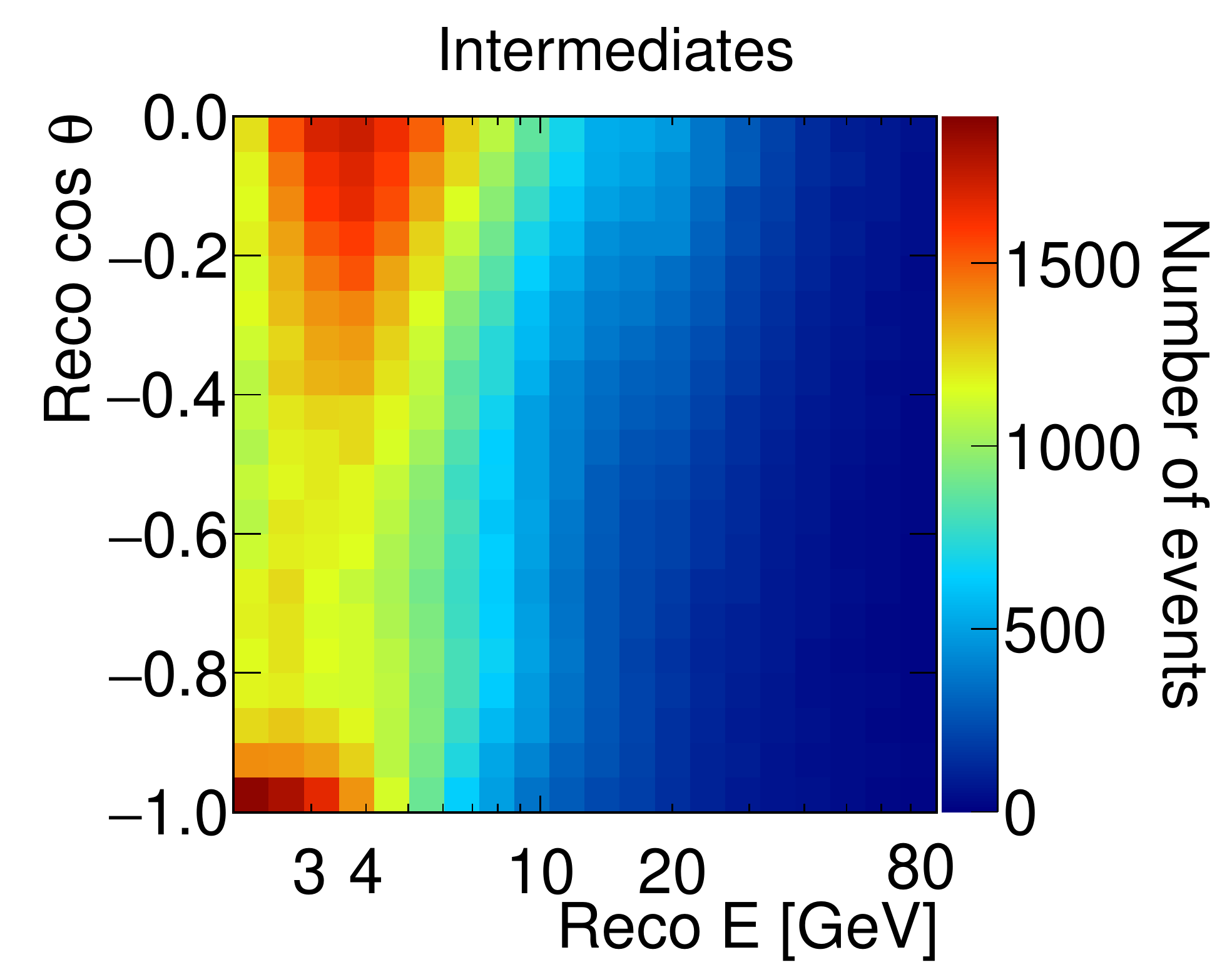}
    \caption{Expected event distribution for ORCA in 3~PID classes for 6~years of exposure and true normal ordering  assumption, with the oscillation parameter values from Ref.~\cite{Esteban:2018azc}.}
    \label{evt_distribution}
\end{figure}

Fig.~\ref{evt_distribution} depicts the expected neutrino event distributions calculated with SWIM for all PID classes, binned in  reconstructed variables ($E_\mathrm{reco}, \cos\theta_\mathrm{reco}$) for 6~years of data taking with ORCA, assuming normal ordering and oscillation parameters from Ref.~\cite{Esteban:2018azc}. In this study, 20 linear bins are chosen for the reconstructed cosine of the zenith angle while the reconstructed energy is binned logarithmically with 20 bins in the range of 2--80~GeV.

\subsection{Sensitivity analysis}\label{ORCAanalysis}

The ORCA analysis uses the 2D distribution of expected neutrino events in each PID class as a function of the  reconstructed neutrino energy and the cosine of the zenith angle. The SWIM package performs the computation and minimization of the test statistic chosen as the  log-likelihood ratio between a hypothetical model and data, which for this analysis is an Asimov dataset with an assumed true hypothesis.

The ORCA $\chi^2$ is built as follows:

\begin{equation}\label{Chi2_Poisson}
    \chi^2 = -2 \sum_{l} \left( d_l - \beta_l\mu_l - d_l \ln \frac{d_l}{\beta_l\mu_l} \right) + \sum _j \frac{(p_j-p^0_j)^2}{\sigma_j ^2} + \sum_k \frac{(1-\beta_k)^2}{\varrho_k^2}.
\end{equation}
The first sum is the Poisson likelihood of the data $d_l$ given the expectation $\mu_l$ at bin $l$, where the latter also  depends on the nuisance parameters $p_j$. The $\beta$ parameters are introduced based on the ``Beeston-Barlow light method"~\cite{Barlow:1993dm} to account for fluctuations due to finite MC statistics.  The second sum accounts for the Gaussian priors on nuisance parameters $p_j$ with mean values $p_j^0$ and variances  $\sigma_j$. The third sum  represents the Gaussian priors to the $\beta$ parameters that are expected to be normally distributed.

The fluctuations $\beta_k$ are assumed to be bin-to-bin uncorrelated and independent of the model parameters. Given this assumption, the values of $\beta_k$ are solved analytically as the solution of $\partial \chi^2/\partial \beta_k = 0$. In addition, the variances $\varrho_k$ of $\beta_k$ can also be evaluated with a probabilistic model which describes the calculation of the response matrix as a single binomial process~\cite{Casadei:2009ic,Paterno:2004cb}. Finally, both $\beta$ and its variance can be estimated analytically and used directly in the calculation of the $\chi^2$ without any requirement for additional minimization. The full description of this procedure can also be found in Ref.~\cite{Bourret:2018kug}. This implementation results in a $\sim 0.2\sigma$ decrease in the sensitivity, reflecting a correction of the overestimation caused by the limited MC sample size.

Two different sets of systematic uncertainties are used in this study (see~Tab.~\ref{syst}). The  ``baseline'' scenario corresponds to the standard set of ORCA systematics adopted for oscillation analyses, similar to Ref.~\cite{Aiello:2021jfn}. Uncertainties related to the incident flux include the spectral index of the neutrino flux energy distribution (free without any constraints) and the flux skew systematics. These skew parameters are introduced to describe the uncertainties in the ratio of different event types, namely $\nu_e/\bar{\nu}_e$, $\nu_\mu/\bar{\nu}_\mu$, and $(\nu_e+\bar{\nu}_e)/(\nu_\mu+\bar{\nu}_\mu)$, while preserving in each case the total associated flux \cite{Bourret:2018kug}. They are constrained with the priors adapted from Ref.~\cite{Barr:2006it}. A NC normalization systematic is implemented as a scaling factor for the number of NC events. To account for detector-related uncertainties, an energy scale systematic is introduced as a global shift of the neutrino true energy in all detector response functions. This implementation captures the effect of undetected variations in the parameters affecting the amount of light recorded by the detector, such as the absorption length and the PMT efficiencies, that would not be accounted for in the reconstruction~\cite{Adrian-Martinez:2016fdl}. Finally, normalization factors for each PID class are also included to account for any possible systematic effects (in the flux, cross-section, or detector response) that would vary the total number of events in each class.

\begin{table}[t!]
    \centering
\caption{Baseline (see Ref.~\cite{Aiello:2021jfn}) and optimistic (see Ref.~\cite{Bezerra:2019dao})  scenarios for the treatment of systematics considered in the ORCA analysis. The cross ($\times$) indicates that the systematic is not included.}
\begin{tabular}{c|c|c}
\label{syst}
 Parameter & Baseline scenario &
 Optimistic scenario \\\hline
 Flux spectral index & \multicolumn{2}{c}{free}\\
 Flux $\nu_e/\bar{\nu}_e$ skew & \multicolumn{2}{c}{7\% prior}\\
Flux $\nu_\mu/\bar{\nu}_\mu$ skew & \multicolumn{2}{c}{5\% prior}\\
Flux $(\nu_e+\bar{\nu}_e)/(\nu_\mu+\bar{\nu}_\mu)$ skew & \multicolumn{2}{c}{2\% prior}\\
NC normalization & \multicolumn{2}{c}{10\% prior}\\
Detector energy scale & 5\% prior & $\times$ \\
PID-class norm. factors &  free & $\times$\\
Effective area scale & $\times$ &10\% prior\\
 Flux energy scale & $\times$ & 10\% prior
\\\hline
\end{tabular}
\end{table}

\begin{figure}[b!]
\vspace{1cm}
\centering
\includegraphics[scale=0.6]{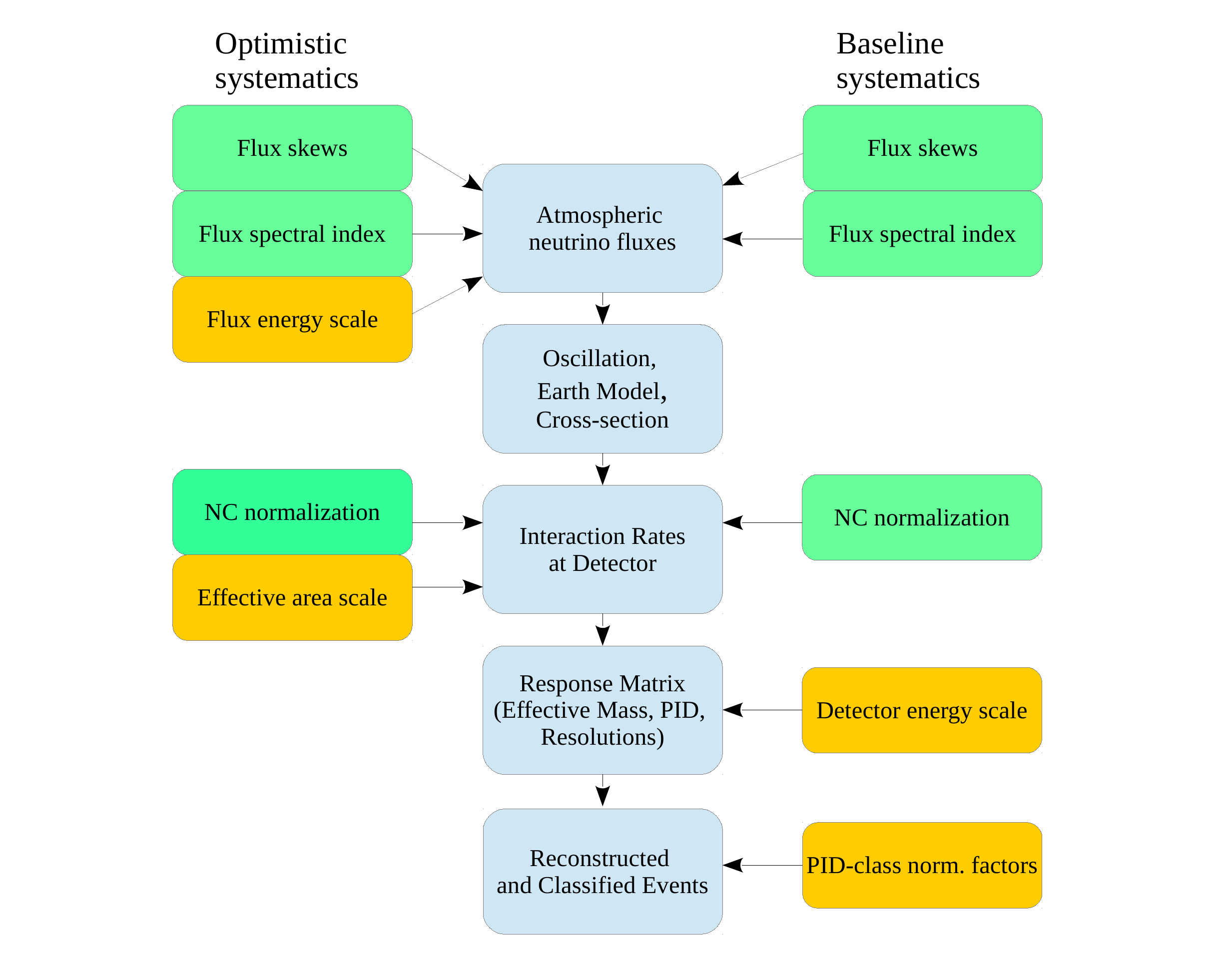}
\caption{Implementation of the two different systematic approaches in the SWIM workflow used in the ORCA analysis.}
\label{fig:orcasens:systchart}
\end{figure}

A second scenario is based on the study of Ref.~\cite{Bezerra:2019dao},  developed  for the PINGU detector. That analysis does not apply normalization factors to the PID classes. It uses an overall scaling factor which represents a universal systematic uncertainty on all effective areas (or equivalently, on the combined $\nu + \bar{\nu}$ event rate). This effective area scaling, together with an energy scale uncertainty introduced at the flux level, are the only systematics introduced to account, {\it e.g.}, for potential variations in the detection efficiency of the optical modules.
Contrary to the baseline case, these systematics are not introduced at the detector response level and are therefore considered as more optimistic in the rest of our study.
The difference between the two approaches is further illustrated in~Fig~\ref{fig:orcasens:systchart}, showing the implementation of each set of systematic uncertainties into the workflow of the SWIM framework. The ``baseline'' systematic set is believed to be more accurate to describe the uncertainties in the ORCA detector. It is therefore used for all presented  results, unless when stated explicitly that the ``optimistic'' systematics are used for the sake of cross-checks and comparisons.

\section{Combined JUNO and ORCA analysis}
\label{sec:ana}

\subsection{Combination strategy}
\label{sec:ana:combination}

As described in Secs.~\ref{sec:juno} and \ref{sec:orca}, the detectors involved in this combined analysis work in very
different conditions. This is true both in terms of their
detection techniques and backgrounds, and of the sources and energies of neutrinos relevant for each individual analysis.  The only
common parameters to both experiments are the neutrino oscillation parameters, which are the
core of the present study.

It is also important to note at this point that not all parameters used to describe standard
neutrino oscillations have an impact on the results of this analysis.
On one hand, JUNO has no sensitivity to either $\theta_{23}$ or $\delta_{\text{CP}}$ as this experiment
measures
$\bar{\nu}_e \rightarrow \bar{\nu}_e$ oscillations which do not depend on those parameters~\cite{An:2015jdp}.
On the other hand,
ORCA has negligible sensitivity to $\theta_{12}$ and $\Delta m^2_{21}$ as the
measured $\nu_\mu + \bar{\nu}_\mu$ oscillations happen at a much smaller $L/E$ than the one
required for the development of oscillations with a frequency given by
$\Delta m^2_{21}$~\cite{Adrian-Martinez:2016fdl}.
The four oscillation parameters  that impact a single experiment are
accounted for implicitly in the $\chi^2$ function computation for JUNO and ORCA following
the prescription outlined below,
while the remaining two oscillation parameters, $\Delta m^2_{31}$ and $\theta_{13}$, have
to be considered explicitly in the joint analysis.

In JUNO, for every value of $\Delta m^2_{31}$ and $\theta_{13}$,
the $\chi^2$ function is minimized using a grid with 61 uniformly spaced values of
$\sin^2 \theta_{12}$ between 0.30225 and 0.31775.
The value of $\Delta m^2_{21}$ is kept fixed at its assumed true value given that JUNO will
be able to determine this parameter quickly.
Studies have shown that
profiling this parameter or keeping it fixed would have an impact smaller than
about 0.1 units of $\chi^2$, which is negligible in the joint analysis.

In ORCA, for every value of $\Delta m^2_{31}$ and $\theta_{13}$,
the $\theta_{12}$ and $\Delta m^2_{21}$ values are kept fixed to their assumed true values given that
ORCA has little sensitivity to those parameters, while $\theta_{23}$ and $\delta_{\text{CP}}$
are minimized without constraints.
This minimization is performed twice, with the initial value of $\theta_{23}$ being located in a
different octant for each minimization.
Only the smallest
value is kept as the global minimum of the $\chi^2$ for ORCA.
This is done to ensure that the minimizer is not trapped in a possible local minimum.

In order to combine the separate JUNO and ORCA analyses, their obtained $\chi^2$ values
at a fixed test value of $\Delta m^2_{31}$ and $\theta_{13}$ are calculated
and summed.
The true value of the oscillation parameters considered in this study are the
best-fit values from Ref.~\cite{Esteban:2018azc} obtained ``with SK data'', unless it is
explicitly stated otherwise.
For added clarity, those parameters are explicitly shown in Tab.~\ref{tab:ana:true_osc}.
Given that neither JUNO nor ORCA are as sensitive to $\theta_{13}$ as current reactor neutrino
experiments~\cite{Adey:2018zwh,Bak:2018ydk,DoubleChooz:2019qbj}, a prior on that parameter was added to the combined
$\chi^2$ from Ref.~\cite{Esteban:2018azc}.
The full expression used is shown in Eq.~\eqref{eq:ana:combined_chi2} where
$\Delta m^2_{31}$ and $\theta_{13}$ are the tested values of those oscillation parameters,
and the last term corresponds to the added prior with
$\sin^2\theta_{13}^{\text{GF}}$ being the current global best fit for $\sin^2 \theta_{13}$
and $\sigma_{\sin^2 \theta_{13}^{\text{GF}}}$ its uncertainty.

\begin{equation}
  \chi^2 \! \left(\Delta m^2_{31}, \theta_{13}\right) =
    \chi^2_{\text{JUNO}} \! \! \left(\Delta m^2_{31}, \theta_{13}\right) +
    \chi^2_{\text{ORCA}} \! \! \left(\Delta m^2_{31}, \theta_{13}\right) +
    \frac{\left(\sin^2 \theta_{13} - \sin^2 \theta_{13}^{\text{GF}}\right)^2}{\sigma^2_{\sin^2 \theta_{13}^{\text{GF}}}}.
  \label{eq:ana:combined_chi2}
\end{equation}

For each set of true parameters studied,
the combined $\chi^2$ from Eq.~\eqref{eq:ana:combined_chi2} is calculated for each
NMO in a $101 \times 21$ grid in the $\left(\Delta m^2_{31}, \sin^2 \theta_{13}\right)$ space,
called the $\chi^2$ map,
centered around the assumed true values of the oscillation parameters
and spanning uniformly
a $\pm 10\%$ interval in $\Delta m^2_{31}$ from the central value and
a $\pm 6\%$ interval in $\sin^2 \theta_{13}$ from the central value.
More explicitly, when assuming true normal ordering with the best-fit values from
Ref.~\cite{Esteban:2018azc}, the tested values of $\Delta m^2_{31}$
in the grid will run from ${-2.78080 \times 10^{-3}}$~eV$^2$ to ${-2.27520 \times 10^{-3}}$~eV$^2$ and
from ${2.27520 \times 10^{-3}}$~eV$^2$ to ${2.78080 \times 10^{-3}}$~eV$^2$
with step of ${0.00506 \times 10^{-3}}$~eV$^2$,
and those of $\sin^2 \theta_{13}$ in the grid will be
from $0.0210278$ to $0.0237122$ with a step of $0.0001342$.
It is worth noting that when the true value of the oscillation parameters
is changed, as in Sec.~\ref{sec:sens:orca}, or when assuming inverted ordering,
the grid described above is changed so that
the central value of the grid corresponds to the true oscillation parameters.

Using the $\chi^2$ map above, for each set of true oscillation parameters tested, the
NMO sensitivity is determined by calculating the
$\overline{\Delta \chi^2} = \chi^2_{WO} - \chi^2_{TO}$, where $\chi^2_{WO}$ ($\chi^2_{TO}$)
is the
minimum value of $\chi^2$ in the $\chi^2$ map in the wrong (true) ordering region of the map.
The $\overline{\Delta \chi^2}$ is then converted into a median sensitivity
$S(\sigma) = \sqrt{\overline{\Delta \chi^2}}$~\cite{Wilks:1938dza}.
The same procedure is also used separately for ORCA and JUNO to obtain the corresponding
non-combined sensitivities, computed for each experiment alone.

The $\overline{\Delta \chi^2}$ notation is used above rather than $\Delta \chi^2$,
to highlight the fact
that an Asimov approximation is being used in the entirety of this paper, therefore,
the median sensitivity is always calculated.

\begin{table}[t!]
  \begin{center}
    \caption{Global best-fit values for the oscillation parameters (from Ref.~\cite{Esteban:2018azc})
      and assumed to be the ``true value'' in this analysis.
      Uncertainties are shown for the parameter where a prior based on the global best-fit value was used.
      }
    \label{tab:ana:true_osc}
    \begin{tabular}{l|cc}
      Parameter & Normal Ordering & Inverted Ordering \\ \hline
      $\sin^2 \theta_{23}$ & 0.563 & 0.565 \\
      $\sin^2 \theta_{13}$ & 0.02237$^{+0.00066}_{-0.00065}$ & 0.02259$\pm$0.00065 \\
      $\Delta m^2_{31}$ & $2.528 \times 10^{-3}$~eV$^2$ & $-2.435 \times 10^{-3}$~eV$^2$ \\
      $\delta_\text{CP}$       & 221$^\circ$ & 282$^\circ$ \\
      $\sin^2 \theta_{12}$ & \multicolumn{2}{|c}{0.310} \\
      $\Delta m^2_{21}$  & \multicolumn{2}{|c}{$7.39 \times 10^{-5}$~eV$^2$} \\
    \end{tabular}
  \end{center}
\end{table}

\subsection{Results}
\label{sec:results}

\begin{figure}[tbh]
  \flushright
  \includegraphics[width=0.95\linewidth]{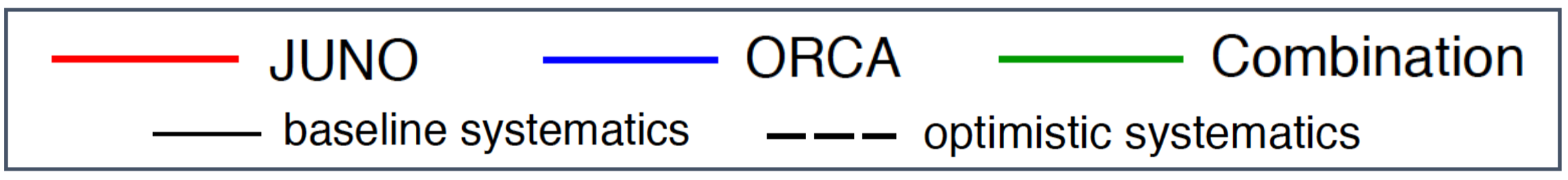} \\
     \vspace*{0.2cm}
  \includegraphics[width=0.49\linewidth]{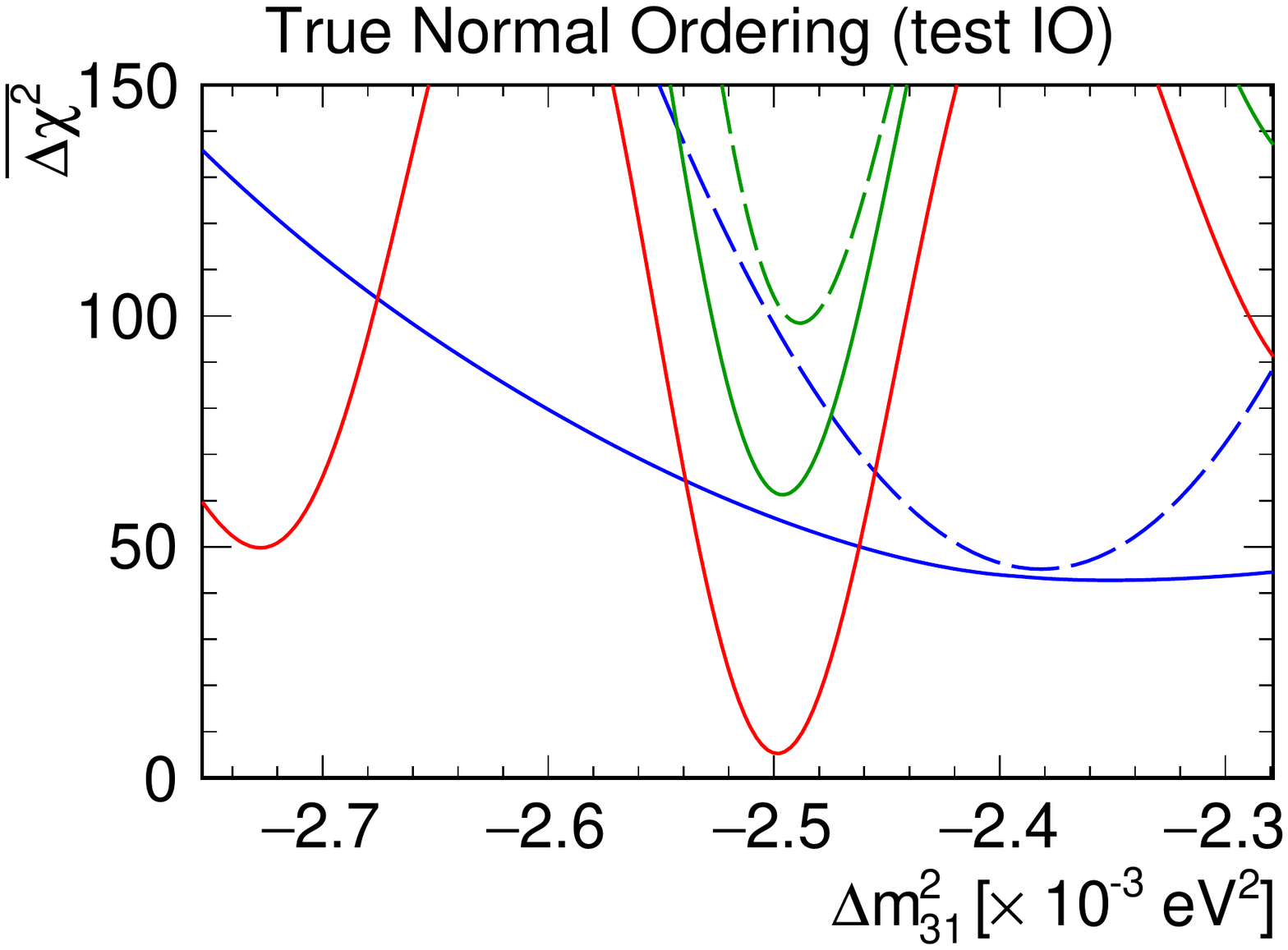}
  \includegraphics[width=0.49\linewidth]{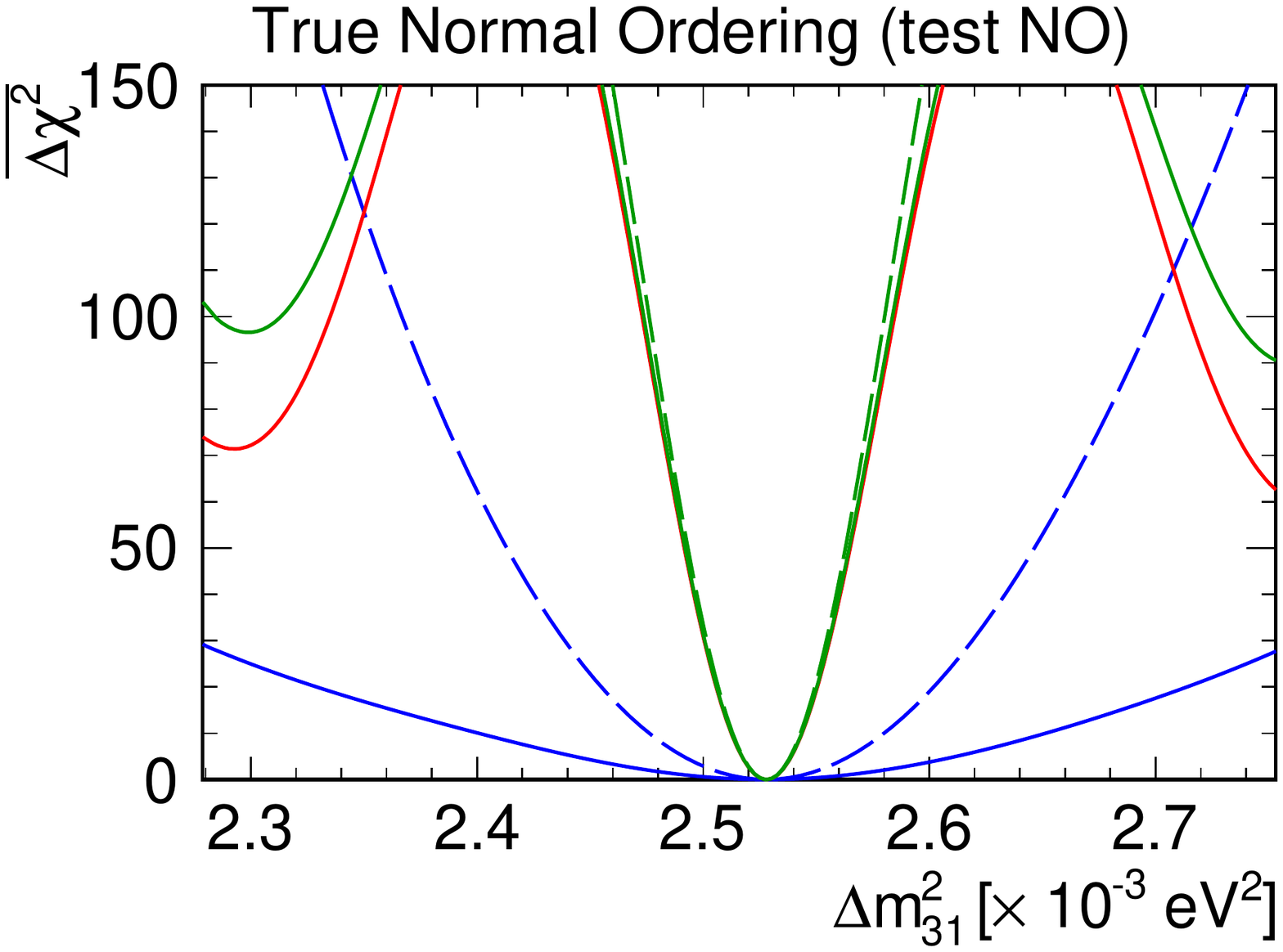}\\
  \includegraphics[width=0.49\linewidth]{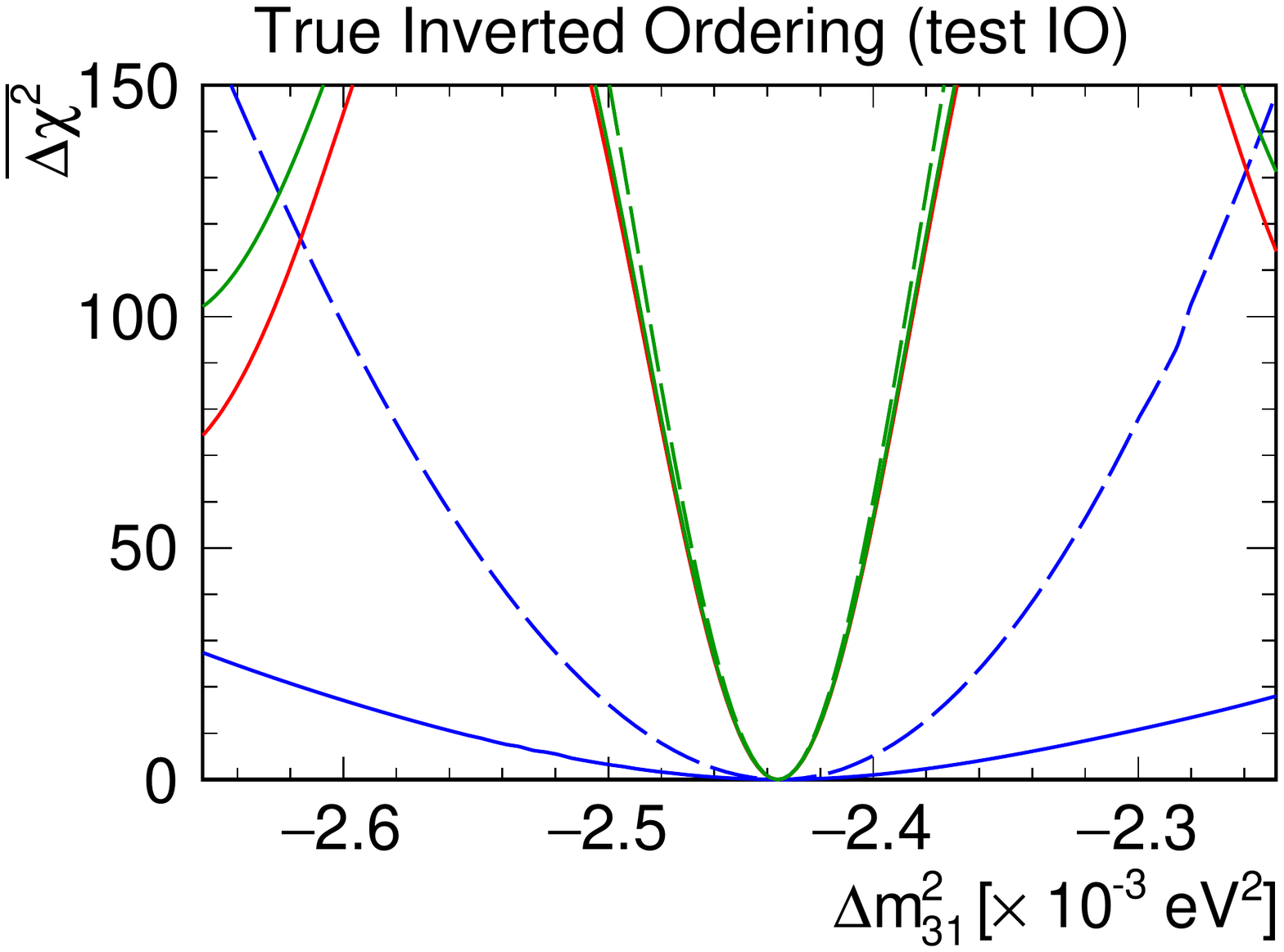}
  \includegraphics[width=0.49\linewidth]{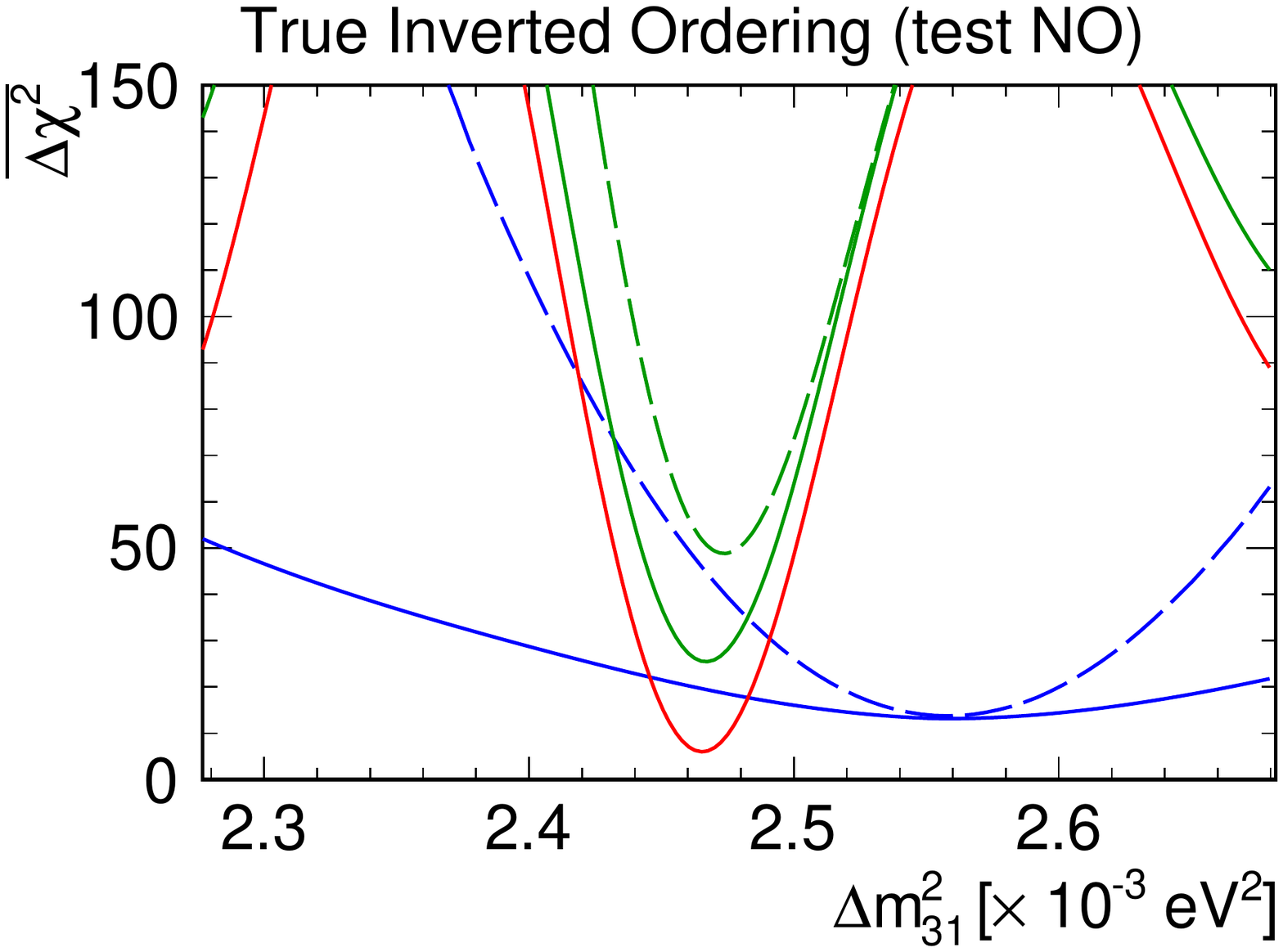}
  \caption{$\overline{\Delta \chi ^2}$ profile for only JUNO (red), only ORCA (blue),
  and the combination of JUNO and ORCA (green) as a function of test values
  of $\Delta m^2_{31}$ for 6 years of data taking assuming
  baseline (solid) or optimistic (dashed) systematics.
  }
  \label{fig:Dm31Scan}
\end{figure}

Fig.~\ref{fig:Dm31Scan} depicts the profile $\overline{\Delta \chi^{2}}$ scan on the test values of $\Delta m_{31}^2$ with 6 years of JUNO and ORCA data taking.
The four profiles correspond to true normal (top) and inverted (bottom) orderings
while fitting the true or wrong  ordering.
Since the Asimov dataset is used, when assuming the true ordering on the fit both experiments show the same best-fit values at the true $\Delta m_{31}^2$ and their $\overline{\Delta \chi^{2}}$ minima yield zero. However, when assuming wrong ordering on the fit, the minima of $\overline{\Delta \chi^{2}}$ are no longer at zero and the sensitivity to the NMO is obtained from this difference. After 6 years, JUNO will be able to exclude the wrong ordering with the significance of $\sim 2.3\sigma$ for either NMO. On the other hand, ORCA is expected to reach a significance of more than  $6\sigma$ ($3\sigma$) for true NO (IO).

Fig.~\ref{fig:Dm31Scan} also shows how the combination of JUNO and ORCA would exceed the NMO sensitivity of each experiment alone. The key advantage of the combination comes from the tension in $\Delta m_{31}^2$ best fits of the two experiments when assuming the wrong ordering.
This tension arises from the fact that each experiment observes neutrino oscillations starting from a different neutrino flavor ($\bar\nu_e$ for JUNO, $\nu_\mu + \bar\nu_\mu$ for ORCA). Due to this difference the effective oscillation frequency will be a different combination of the various $\Delta m^2_{ij}$ for each experiment~\cite{Nunokawa:2005nx,deGouvea:2005hk}.
Since the combination requires a single resulting $\Delta m_{31}^2$ best fit, this tension together with strong constraints in $\Delta m_{31}^2$ from both experiments, and particularly from JUNO, provides the synergy effect in which the combined $\overline{\Delta \chi^2}$ minimum is enhanced to a higher value than simply adding the $\overline{\Delta \chi^2}$ minima from each experiment.
This latter scenario, in which the median sensitivity can be obtained as the square root of the sum, will be referred to as ``simple sum'' in the following discussion. It is
shown only to highlight the benefit from doing the combination between JUNO and ORCA properly.

In Tab.~\ref{table:NMO6years}, the NMO sensitivities after 6 years of collected data are presented for the combination, each experiment standalone, and the ``simple sum'' of their sensitivities. The combination reaches $8\sigma$ for true NO and $5\sigma$ for true IO. This combined sensitivity exceeds the ``simple sum'' case, which only obtains $7\sigma$ for true NO and $4\sigma$ for true IO. More importantly, a $5\sigma$ significance is achieved for both NMO scenarios within 6 years of combined analysis while each experiment alone, or the ``simple sum'' of sensitivities, cannot achieve the same performance, sometimes even at significantly longer timescales.

\begin{table}
\centering
\caption{Asimov median sensitivity to NMO after 6 years of data taking for each experiment alone, the ``simple sum'', and the combination of the two experiments, assuming the baseline scenario for systematics.}
\label{table:NMO6years}
\begin{tabular}{c|c|c|c|c}
True NMO  & JUNO, 8 cores&ORCA &Simple Sum&Combination\\\hline
NO   & $2.3\sigma$ & $6.5\sigma$ & $6.9\sigma$ &$7.8\sigma$\\
IO   & $2.4\sigma$ & $3.6\sigma$ & $4.3\sigma$ &$5.1\sigma$
\\
\end{tabular}
\end{table}

\begin{figure}[t]
\flushright
  \includegraphics[width=0.95\linewidth]{Legend_Figs/Legend_systematicsORCA.png} \\
     \vspace*{0.2cm}
  \includegraphics[width=0.49\linewidth]{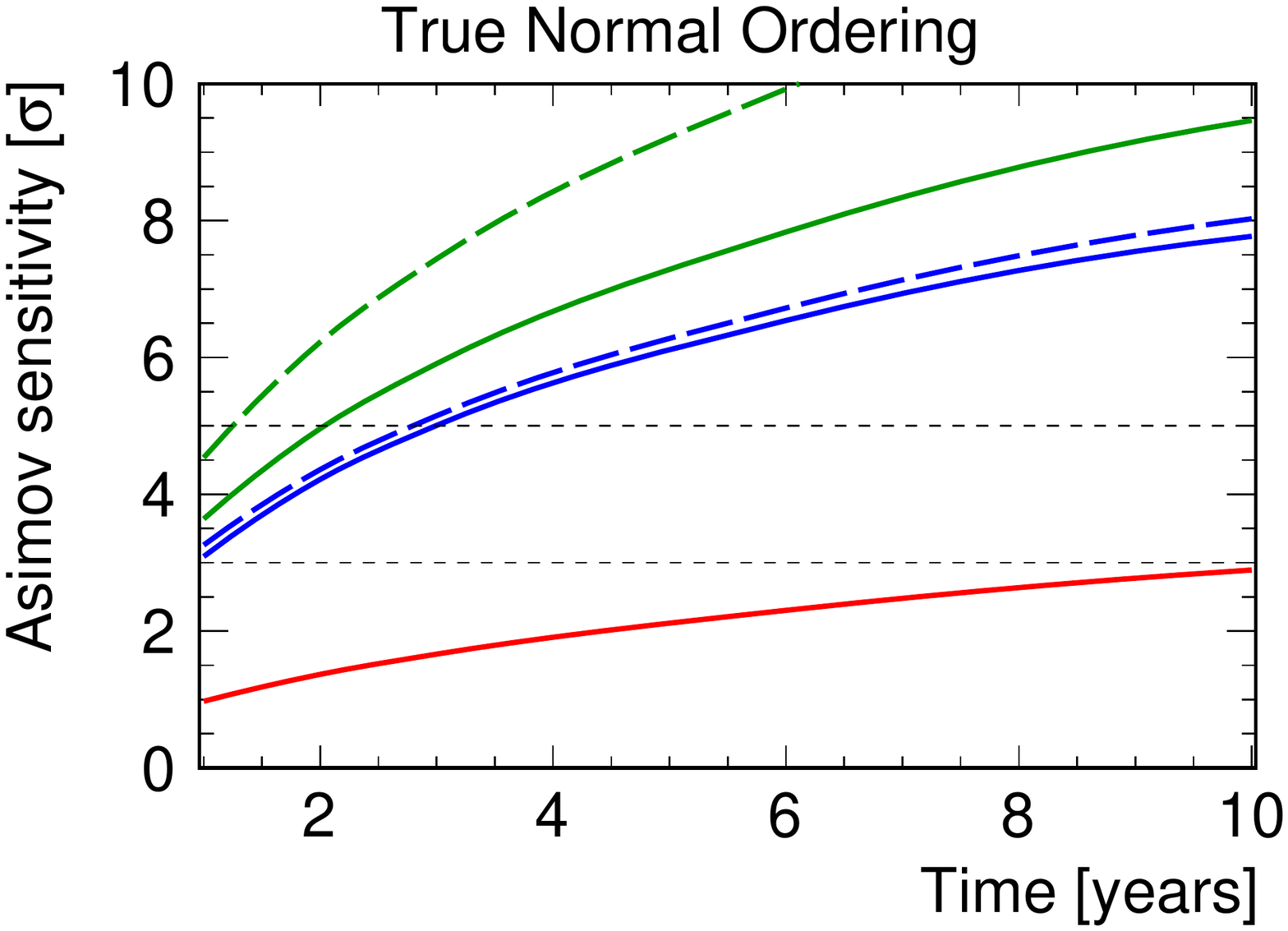}
  \includegraphics[width=0.49\linewidth]{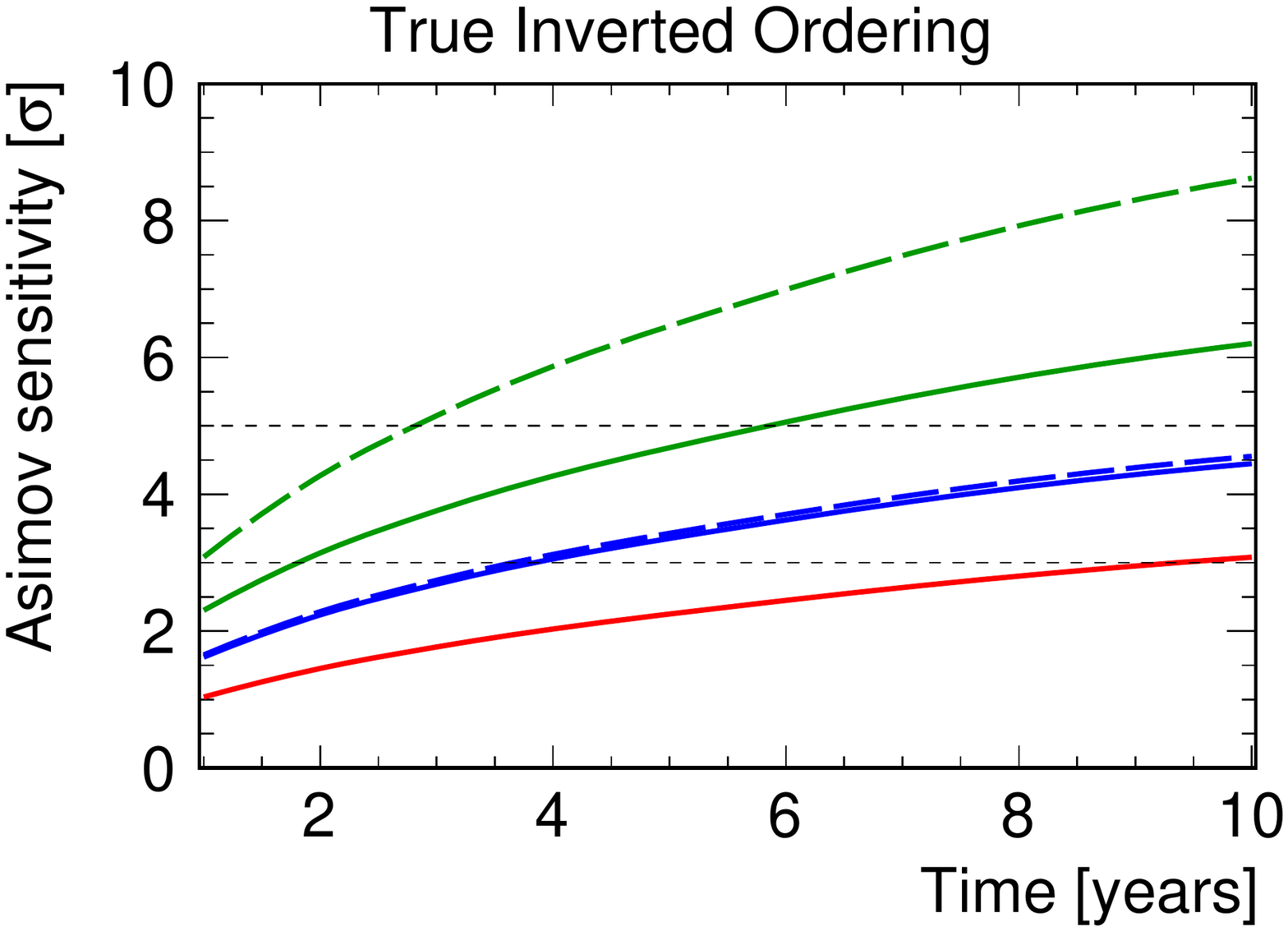}
  \caption{NMO sensitivity as a function of time for only JUNO (red), only ORCA (blue),
  and the combination of JUNO and ORCA (green), assuming baseline (solid) or
  optimistic (dashed) systematics.}
  \label{fig:NMO_Time}
\end{figure}

The time evolution of the NMO sensitivity for JUNO, ORCA, and their corresponding combination is presented in Fig.~\ref{fig:NMO_Time}
 assuming that the two experiments start at the same time. JUNO alone would need 6--10 years of operation to reach 3$\sigma$ of NMO sensitivity. ORCA has the capability to reach a  $5\sigma$ significance after 3~years in the case of true NO.  However, it would take more than 10~years of exposure to reach $5\sigma$ sensitivity in the case of IO.  Due to the synergy effect discussed above, the combination would help significantly to reduce the time needed to reach a $5\sigma$ NMO  sensitivity %by at least a year
 when compared to ORCA, especially if the neutrino mass ordering is inverted. With the combined analysis, a $5\sigma$ significance can be obtained within 2 (6) years in the case of true NO (IO) respectively.

As discussed in Sec.~\ref{ORCAanalysis}, the ORCA analysis is also performed using a set of systematics similar to those of Ref.~\cite{Bezerra:2019dao}, as a cross-check for an optimistic approach. Fig.~\ref{fig:Dm31Scan} and Fig.~\ref{fig:NMO_Time} show that both the optimistic and the baseline systematics give a very similar $\Delta \chi^2$ minimum value and thus yield the same NMO sensitivity for the ORCA-only analysis. However, the optimistic approach provides a much tighter constraint on $\Delta m_{31}^2$, as shown in Fig.~\ref{fig:Dm31Scan}, which causes the combination to reach sensitivities that are 1--2$\sigma$ higher than in the case of the baseline scenario. This comes from the difference in the implementation of the energy scale systematics. The energy scale applied at the detector response (baseline) is more strongly correlated with $\Delta m_{31}^2$ compared to the energy scale at the unoscillated flux (optimistic).

% vim: spelllang=en

\section{Further sensitivity studies}
\label{sec:sens}

\subsection{Impact of energy resolution in JUNO and 10 reactor cores scenario}
\label{sec:sens:juno}

One of the most challenging design specifications of JUNO is the required
energy resolution of the central detector.
Reaching a level of about $3\%/\sqrt{E/\text{MeV}}$ is essential for JUNO to be able to reach
a $3\sigma$ sensitivity to determine the neutrino mass ordering by itself.
In this sense, if the energy resolution worsens to
$3.5\%/\sqrt{E/\text{MeV}}$, the required time to reach a $3\sigma$ sensitivity would increase by
more than a factor of 2~\cite{An:2015jdp}.
A significant amount of effort has been made within the JUNO Collaboration to reach
this goal of $3\%/\sqrt{E/\text{MeV}}$, and a description of how to get there
using a data-driven approach relying on calibration data has been
presented in Ref.~\cite{Abusleme:2020lur},
where a 3.02$\%/\sqrt{E/\text{MeV}}$ energy resolution has been achieved, with a
worsening of this energy resolution to 3.12$\%/\sqrt{E/\text{MeV}}$ after considering
some imperfections in the detector.
Nevertheless, it is still extremely interesting to evaluate the sensitivity of the combined
NMO analysis to the energy resolution of JUNO.

\begin{figure}[t]
  \begin{center}
  \flushright
  \includegraphics[width=0.95\linewidth]{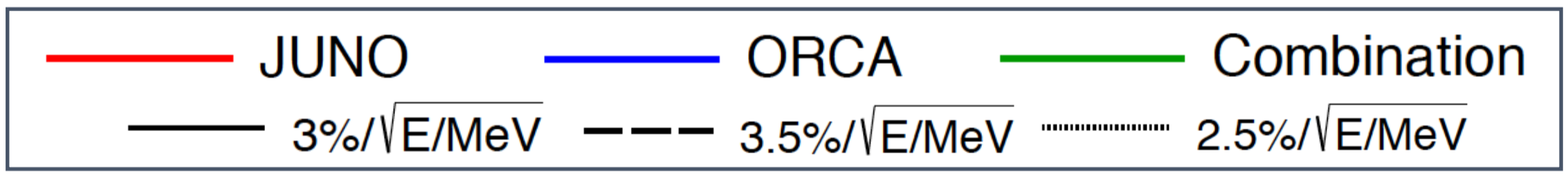} \\
     \vspace*{0.2cm}
    \includegraphics[width=0.49\linewidth]{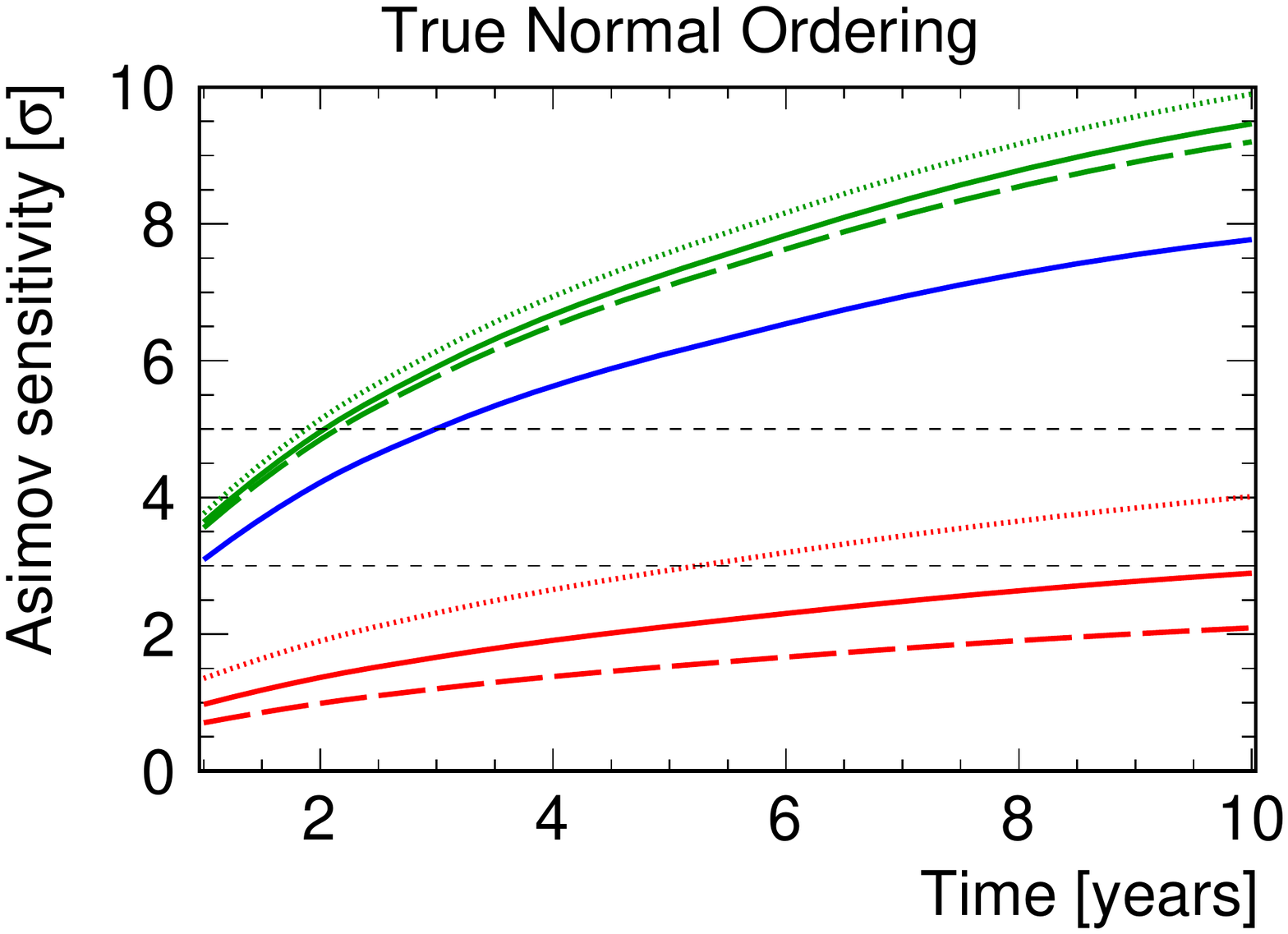}
    \includegraphics[width=0.49\linewidth]{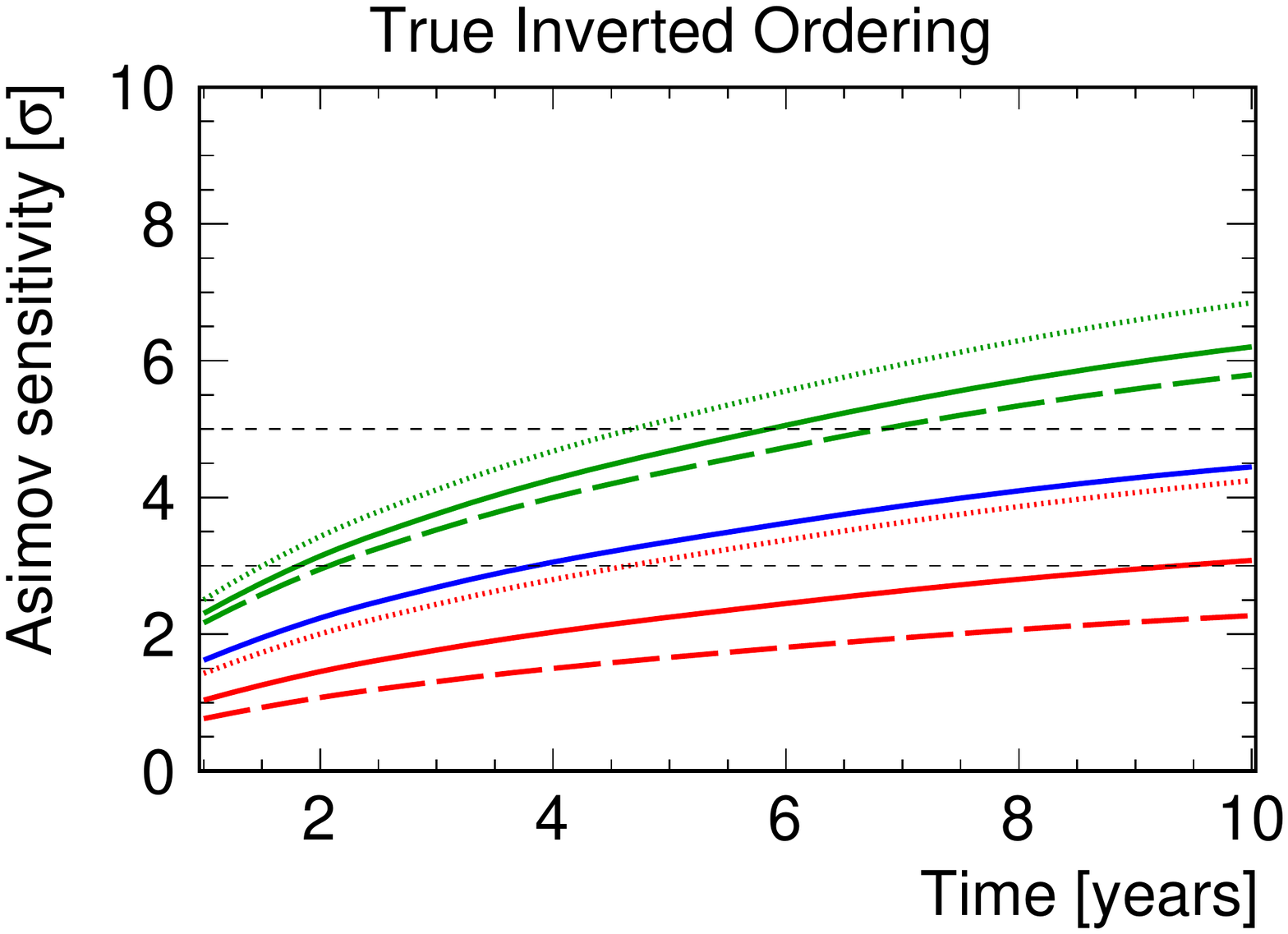}
    \caption{NMO sensitivity as a function of time for only JUNO (red), only ORCA (blue),
    and the combination of JUNO and ORCA (green), considering a better (dotted) and worse (dashed)
    energy resolution for JUNO than the nominal one (solid) by $\pm 0.5\%/\sqrt{E/\text{MeV}}$.}
    \label{fig:sens:juno:energy_resolution}
  \end{center}
\end{figure}

In the present study a $\pm 0.5\%/\sqrt{E/\text{MeV}}$ variation of the energy resolution was
considered.
While this accounts for a larger departure from the JUNO target energy resolution than the one
described above, it was chosen to test the robustness of the combination procedure.
As shown in Fig.~\ref{fig:sens:juno:energy_resolution},
the impact of this variation of the energy resolution in the combined analysis is fairly small
in comparison to the impact on the JUNO-only analysis.
The reason for this small impact is that the added power to discriminate
the neutrino mass orderings in this combination
comes mostly from the displacement between the $\Delta m^2_{31}$ best-fit values obtained by ORCA and JUNO
for the wrong ordering assumption rather than from the direct measurement of the neutrino mass
ordering in JUNO, as discussed previously.
In this scenario, a worsening of the energy resolution would slightly reduce the precision
of JUNO to measure $\Delta m^2_{31}$, while the best-fit value of $\Delta m^2_{31}$
for each ordering would not change significantly. Therefore,
the tension of the $\Delta m^2_{31}$ best fit between JUNO and ORCA remains,
which preserves the high sensitivity of the analysis.

\begin{figure}[ht!]
\vspace{1cm}
\flushright
  \includegraphics[width=0.95\linewidth]{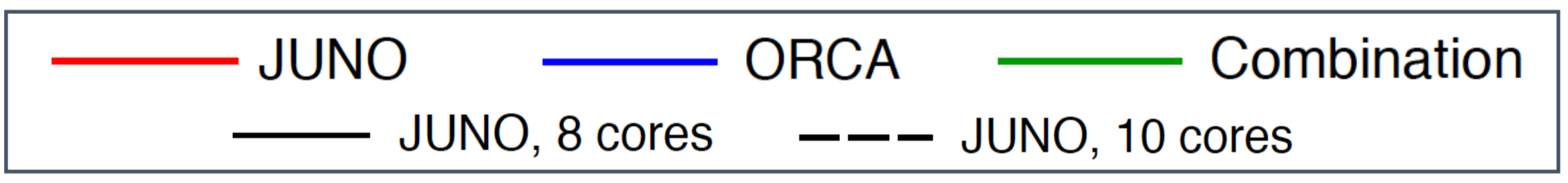} \\
     \vspace*{0.05cm}
  \begin{center}
    \includegraphics[width=0.49\linewidth]{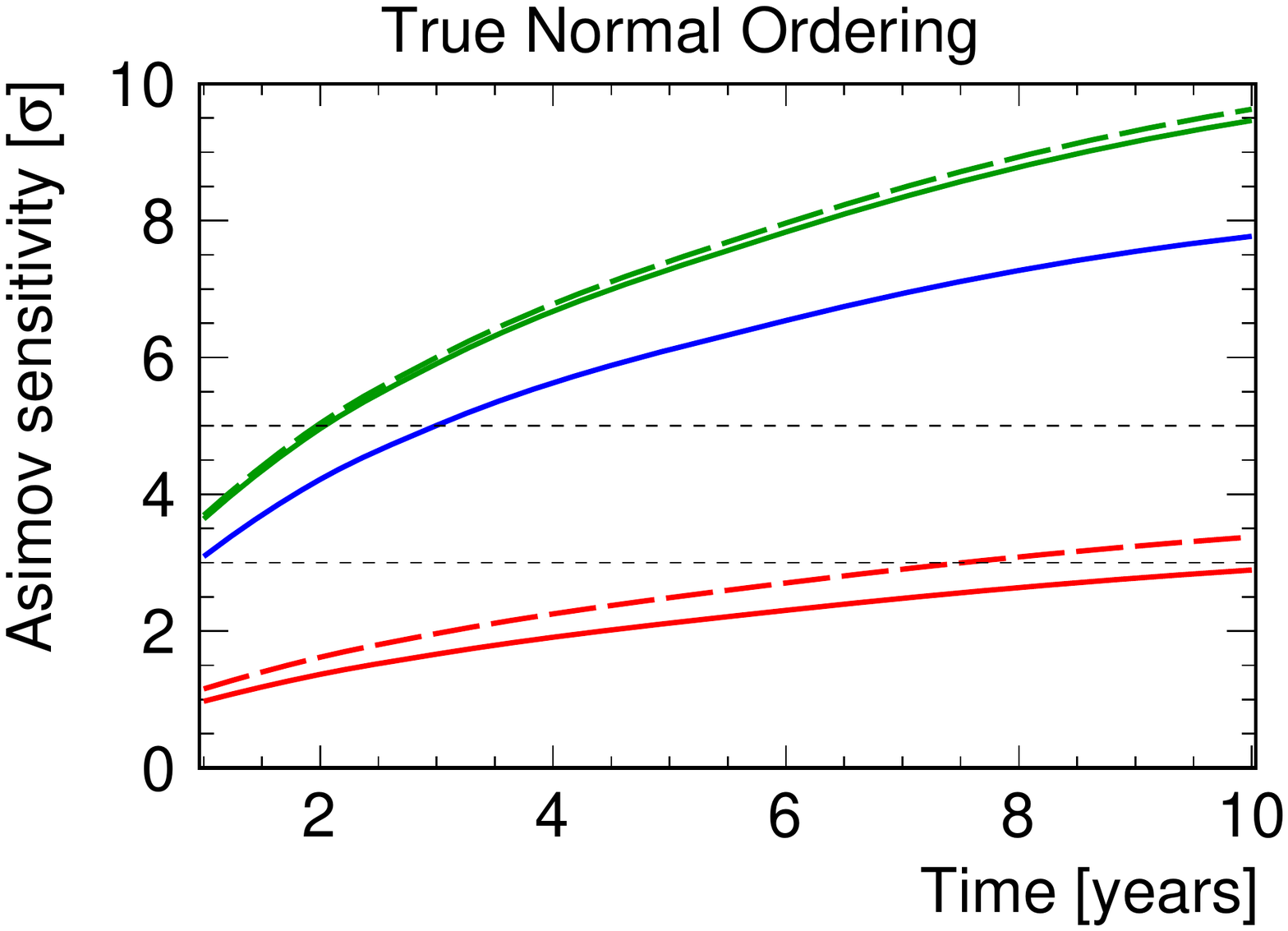}
    \includegraphics[width=0.49\linewidth]{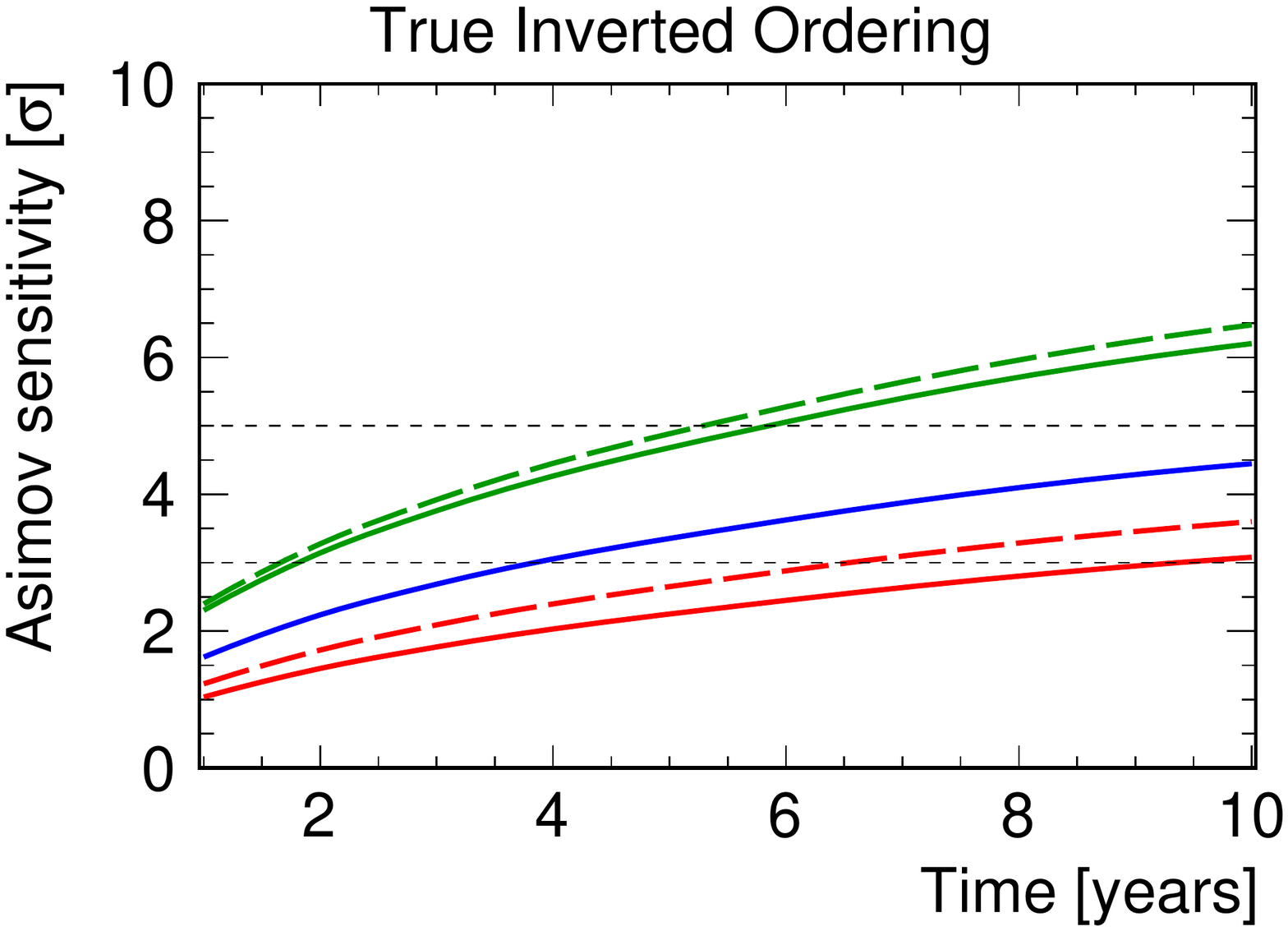}
    \caption{NMO sensitivity as a function of time for only JUNO (red), only ORCA (blue),
    and the combination of JUNO and ORCA (green), considering 2 (solid) or 4 (dashed)
      Taishan NPP reactors, corresponding respectively to 8 or 10~reactor cores at 53~km from JUNO.}
    \label{fig:sens:juno:10cores}
  \end{center}
\end{figure}

As discussed in Sec.~\ref{sec:juno:modeling}, there is a possibility that
2 additional reactors could be built at the Taishan NPP, as originally planned.
This would double the number of neutrinos produced by that NPP.
In this scenario,
JUNO by itself would be able to reach $3\sigma$ about 3~years earlier,
as shown in Fig.~\ref{fig:sens:juno:10cores}.
In combination with ORCA however, there is negligible impact to the time required
for the combined sensitivity to reach $5\sigma$ assuming true normal ordering, at the current
best fit value. About 9~months are gained in the inverted ordering scenario
which is still a significantly smaller impact than for the standalone JUNO.
This behavior is, as in the case of the JUNO energy resolution dependency,
due to the fact that the boost obtained from the combination relies on the difference between the JUNO
and ORCA best-fit values for $\Delta m^2_{31}$ in the wrong ordering scenario,
rather than due to the precision of each experiment to measure the
neutrino mass ordering separately.

\subsection{Dependence on $\Delta m^{2}_{31}$ and $\theta_{23}$}
\label{sec:sens:orca}

This section presents the dependence of the analysis on the true value of the oscillation parameters,
focusing particularly on $\theta_{23}$ and $\Delta m^2_{31}$.
Those parameters are chosen because the true value of $\theta_{23}$ is known to have a strong influence
on the ORCA sensitivity,
and because the boost in sensitivity in the combined analysis, as discussed previously, is directly
tied to the $\Delta m^2_{31}$ measurement, and therefore it is essential to ensure that such a boost is
valid for any true value of $\Delta m^2_{31}$.

\begin{figure}[ht!]
\flushright
  \includegraphics[width=0.95\linewidth]{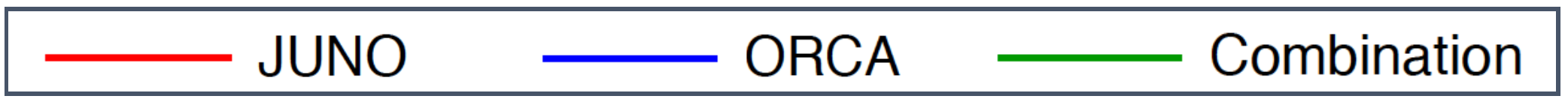} \\
  \vspace*{0.2cm}
\centering
\includegraphics[width=0.49\linewidth]{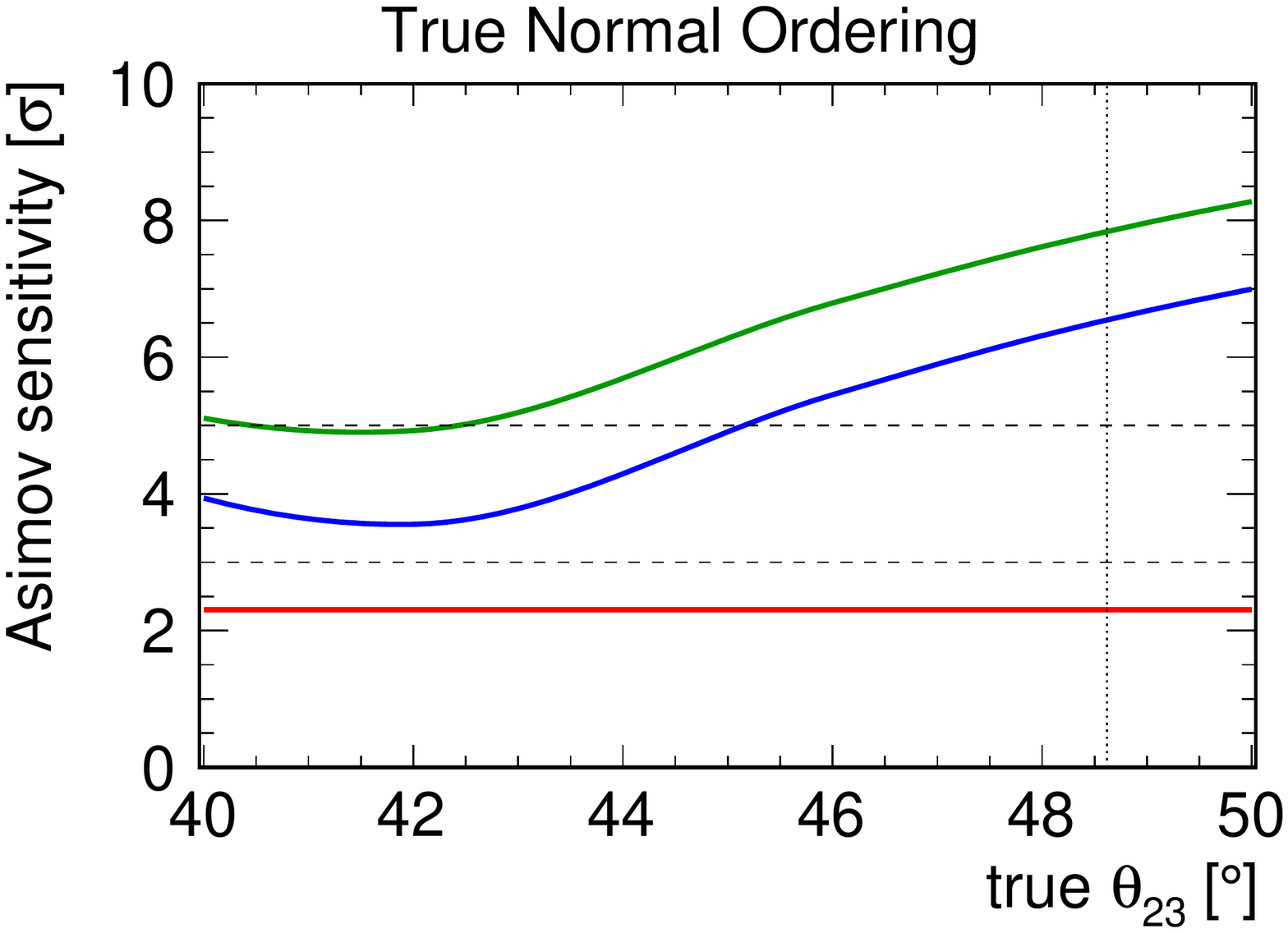}
\includegraphics[width=0.49\linewidth]{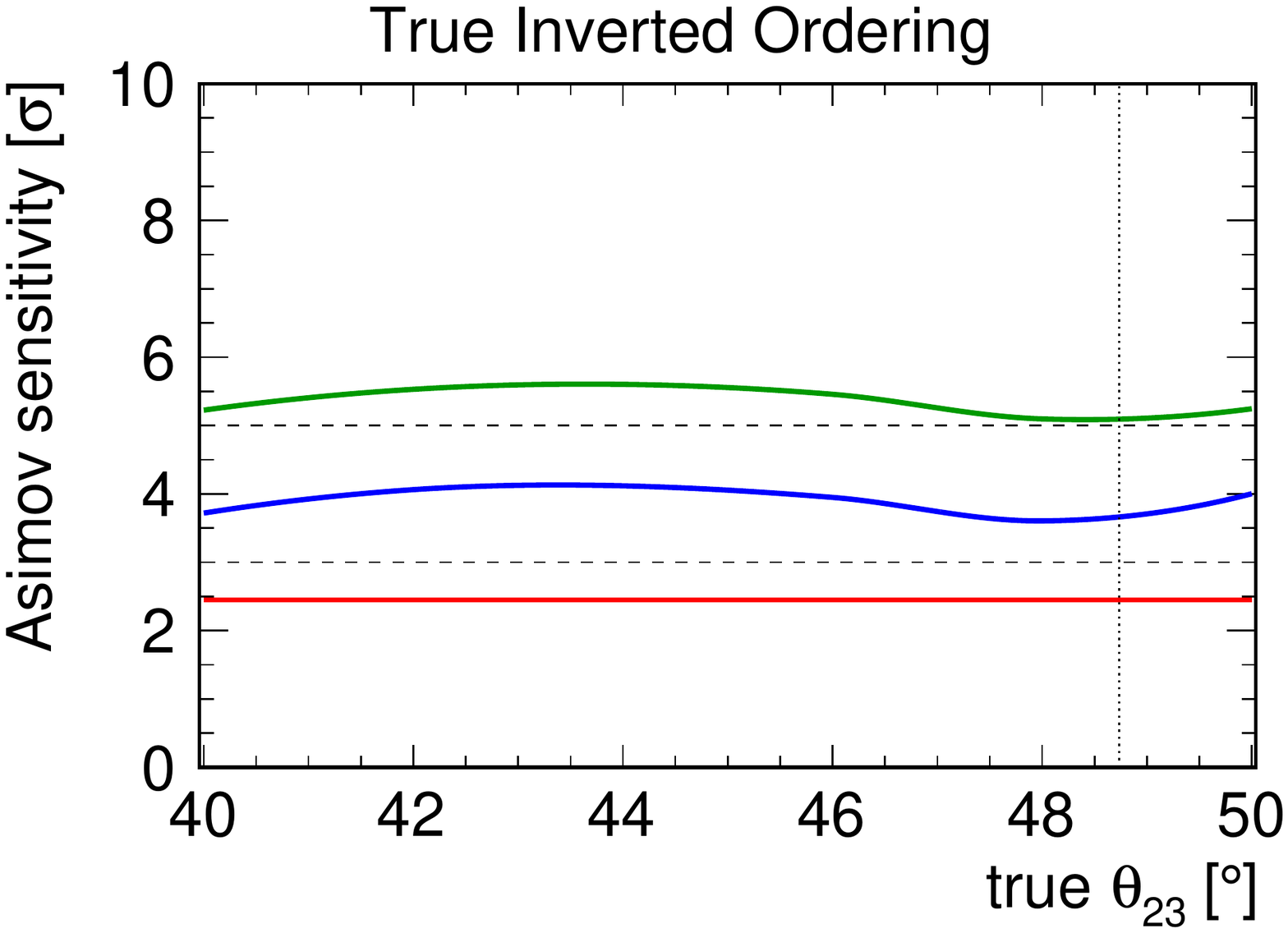}
\caption{NMO sensitivity as a function of the true $\theta_{23}$ value for 6~years of data taking
  for only JUNO (red), only ORCA (blue),
  and the combination of JUNO and ORCA (green).
  The vertical lines indicate the global best-fit values used in this analysis (from Ref.~\cite{Esteban:2018azc}).
  }\label{fig:Th23}
\end{figure}

Fig.~\ref{fig:Th23} shows the dependence of the NMO sensitivity on the true value of $\theta_{23}$ for 6~years of data taking. As mentioned in Sec.~\ref{sec:ana:combination}, JUNO has no sensitivity to $\theta_{23}$. In the case of ORCA however, the NMO sensitivity potential depends strongly on the true value of $\theta_{23}$ as this parameter affects the amplitude of the detected oscillation pattern. After 6~years of data taking, ORCA has the sensitivity to reject the wrong ordering with a significance of 3--7$\sigma$ and only reaches a $5\sigma$ sensitivity for true NO with $\theta_{23}$ in the second octant. The combination curve also follows a similar $\theta_{23}$ dependence as the ORCA-standalone curve, however thanks to the boost from JUNO, it is shifted to higher sensitivities and the joint fit ensures a $5\sigma$ discovery after about 6~years regardless of the true value of $\theta_{23}$ and of the true NMO.

It is worth noting here that the current global best-fit value of $\theta_{23}$ is in the upper octant with values of about $49^\circ$ for both orderings. This value is the one used in the studies described in  Secs.~\ref{sec:results} and \ref{sec:sens:juno}, which explains why in those studies the sensitivity for true NO is always much higher than $5\sigma$ after 6 years of data taking.

\begin{figure}[ht!]
\flushright
  \includegraphics[width=0.95\linewidth]{Legend_Figs/Legend_simple.png} \\
    \vspace*{0.2cm}
\centering
\includegraphics[width=0.49\linewidth]{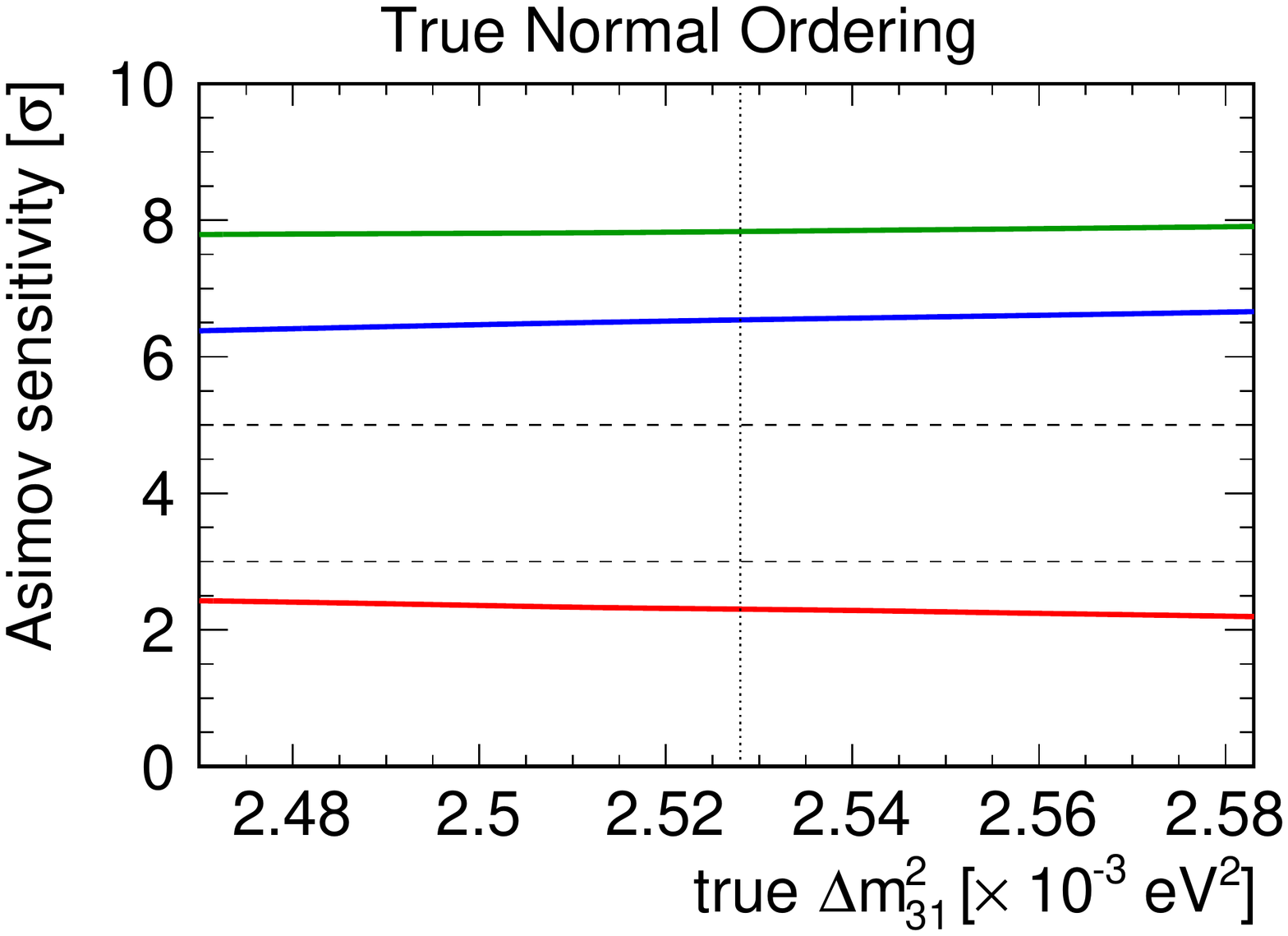}
\includegraphics[width=0.49\linewidth]{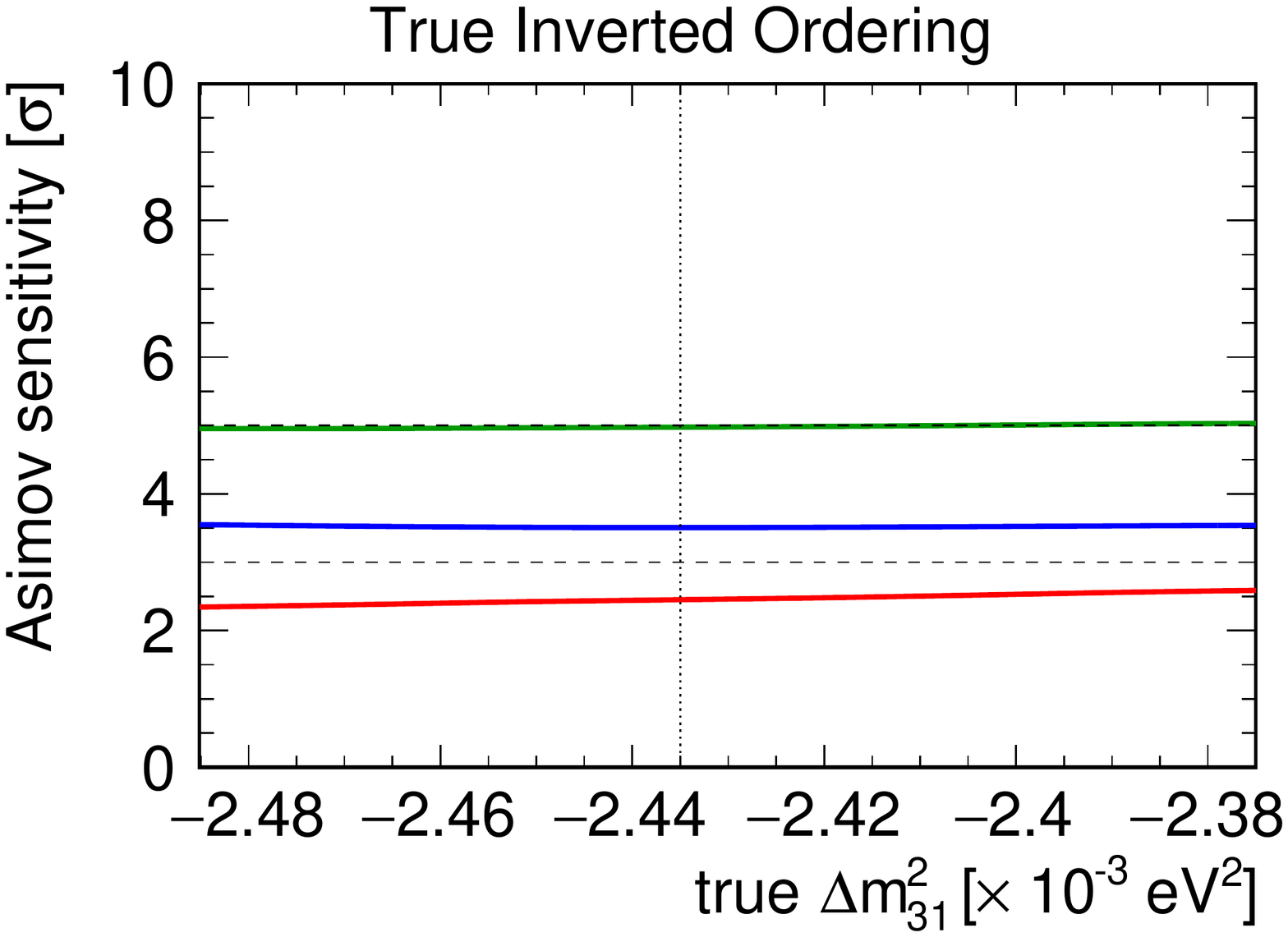}
\caption{NMO sensitivity as a function of the true $\Delta m_{31}^2$ value for 6~years of data taking
  for only JUNO (red), only ORCA (blue),
  and the combination of JUNO and ORCA (green).
  The vertical lines indicate the global best-fit values used in this analysis (from Ref.~\cite{Esteban:2018azc}).
  }\label{fig:Dm31}
\end{figure}

Fig.~\ref{fig:Dm31} illustrates the dependence of NMO sensitivity on the true value of $\Delta m_{31}^2$. Both JUNO and ORCA standalone sensitivities depict a slight dependence on the true value of $\Delta m_{31}^2$ with the opposite slope for each experiment. The combination is also quite flat with respect to the true $\Delta m_{31}^2$, reaching a significance of $8\sigma$ in the case of NO and $5\sigma$ in the case of IO. The effect of the boost described previously, relying on the difference between the wrong ordering measurement of $\Delta m_{31}^2$, is preserved over the whole $\Delta m_{31}^2$ range.

% vim: spelllang=en

\section{Conclusions}
\label{sec:end}

This paper presents an evaluation of the sensitivity to the neutrino mass ordering achieved by a combined analysis
of the JUNO and KM3NeT/ORCA experiments. It is worth pointing out explicitly that in all cases the combined analysis is more powerful
than simply adding the sensitivities for both experiments together.
As discussed above,
this is due to the tension that arises in the $\Delta m^2_{31}$ measurement between
JUNO and ORCA when the wrong neutrino mass ordering is assumed.

The results show that
this combination significantly reduces the time required to reach a $5\sigma$ determination
of the neutrino mass ordering for any value of the oscillation parameters.
In all cases, a
$5\sigma$ measurement can be obtained within 6~years for the combined analysis, while it
could take more than 10~years using only ORCA data, depending on the true ordering.

The gain in time is larger in cases where ORCA alone would
require a longer time for reaching a $5\sigma$ sensitivity due to the
uncertainty on the $\theta_{23}$ value.
In the favorable case of true normal ordering and $\theta_{23}$ in the second octant,
a $5\sigma$ NMO determination would be feasible
after less than 2~years of data taking with the combined analysis.
In this favorable scenario,
which also corresponds to the current global best-fit value,
the neutrino mass ordering would be determined
at least a year ahead of what can be done
using only ORCA data.

The boost for the NMO sensitivity obtained by combining JUNO and ORCA presented in this study is in line with what has been presented by previous studies considering the combination of JUNO with the IceCube Upgrade or with PINGU in Refs.~\cite{Blennow:2013vta,Bezerra:2019dao}.
However, given the differences between PINGU and ORCA, it is important to confirm the result also for the combination of JUNO and ORCA.
Of particular interest is the different treatment of the energy scale systematics between this and previous studies. This uncertainty impacts directly the $\Delta m^2_{31}$ determination with ORCA and thus also the combined result.
As shown in this paper, changing the treatment of this systematic uncertainty
from an optimistic to a more realistic scenario  may significantly affect the power of the combination
of JUNO and ORCA.
Nevertheless, a $5\sigma$ determination of the
neutrino mass ordering can be effectively reached even in the ORCA baseline scenario for systematics.

Because the gain in time to reach the determination of the neutrino mass ordering in
the combination of JUNO and ORCA does not come exclusively from each experiment's own ability
to determine the neutrino mass ordering,
the combination is sensitive to systematic uncertainties and detector
effects in a different way than either experiment do independently.
For instance, even if the JUNO energy resolution is critical for the measurement of the neutrino
mass ordering using only JUNO data, it has only a small impact in the combined analysis.
Alternatively, changing the ORCA systematics between optimistic and baseline systematics
has a small impact on the power of ORCA alone to determine the neutrino mass ordering, however it
has a larger impact on the combined analysis.
These differences arise from the fact that the combination depends strongly on the
measurement of $\Delta m^2_{31}$ by each experiment rather than simply on their measurements of the
neutrino mass ordering directly.
In the cases where the gain in time to reach $5\sigma$ is of only a few years, this
JUNO-ORCA combination is
particularly interesting to provide a somewhat independent validation of the result obtained
by the ORCA experiment alone, with a different dependency on the systematic uncertainties.

% vim: spelllang=en

\acknowledgments

The authors acknowledge the financial support of the funding agencies:
Agence Nationale de la Recherche (contract ANR-15-CE31-0020),
Centre National de la Recherche Scientifique (CNRS),
Commission Europ\'eenne (FEDER fund and Marie Curie Program),
Institut Universitaire de France (IUF),
LabEx UnivEarthS (ANR-10-LABX-0023 and ANR-18-IDEX-0001),
Shota Rustaveli National Science Foundation of Georgia (SRNSFG, FR-18-1268),
Georgia;
Deutsche Forschungsgemeinschaft (DFG),
Germany;
The General Secretariat of Research and Technology (GSRT),
Greece;
Istituto Nazionale di Fisica Nucleare (INFN),
Ministero dell'Univer\-si\-t\`a e della Ricerca (MIUR),
PRIN 2017 program (Grant NAT-NET 2017W4HA7S)
Italy;
Ministry of Higher Education Scientific Research and Professional Training,
ICTP through Grant AF-13,
Morocco;
Nederlandse organisatie voor Wetenschappelijk Onderzoek (NWO),
the Netherlands;
The National Science Centre, Poland (2015/18/E/ST2/00758);
National Authority for Scientific Research (ANCS),
Romania;
Ministerio de Ciencia, Innovaci\'{o}n, Investigaci\'{o}n y Universidades (MCIU): Programa Estatal de Generaci\'{o}n de Conocimiento (refs. PGC2018-096663-B-C41, -A-C42, -B-C43, -B-C44) (MCIU/FEDER), Generalitat Valenciana: Prometeo (PROMETEO/2020/019), Grisol\'{i}a (ref. GRISOLIA/2018/119) and GenT (refs. CIDEGENT/2018/034, /2019/043, /2020/049) programs, Junta de Andaluc\'{i}a (ref. A-FQM-053-UGR18), La Caixa Foundation (ref. LCF/BQ/IN17/11620019), EU: MSC program (ref. 101025085), Spain.

\bibliographystyle{unsrt_modjp}
\bibliography{sample}
\end{document}